# 30 Years of Software Refactoring Research: A Systematic Literature Review

Chaima Abid, Vahid Alizadeh,Marouane Kessentini, Thiago do Nascimento Ferreira and Danny Dig

**Abstract**—Due to the growing complexity of software systems, there has been a dramatic increase and industry demand for tools and techniques on software refactoring in the last ten years, defined traditionally as a set of program transformations intended to improve the system design while preserving the behavior. Refactoring studies are expanded beyond code-level restructuring to be applied at different levels (architecture, model, requirements, etc.), adopted in many domains beyond the object-oriented paradigm (cloud computing, mobile, web, etc.), used in industrial settings and considered objectives beyond improving the design to include other non-functional requirements (e.g., improve performance, security, etc.). Thus, challenges to be addressed by refactoring work are, nowadays, beyond code transformation to include, but not limited to, scheduling the opportune time to carry refactoring, recommendations of specific refactoring activities, detection of refactoring opportunities, and testing the correctness of applied refactorings. Therefore, the refactoring research efforts are fragmented over several research communities, various domains, and objectives. To structure the field and existing research results, this paper provides a systematic literature review and analyzes the results of 3183 research papers on refactoring covering the last three decades to offer the most scalable and comprehensive literature review of existing refactoring research studies. Based on this survey, we created a taxonomy to classify the existing research, identified research trends, and highlighted gaps in the literature and avenues for further research.

## 1 Introduction

For decades, code refactoring has been applied in informal ways before it was introduced and properly defined in academic work. The first known use of the term *Refactoring* in the published literature was in an article written by William Opdyke and Ralph Johnson in September 1990 [1]. William Griswold's Ph.D. dissertation [2], published in 1991, is also one of the first major academic works on refactoring functional and procedural programs. The author defined a set of automatable transformations and described their impact on the code structure. One year later, William Opdyke also published his Ph.D. dissertation [3] on the Refactoring of object-oriented programs. In 1999, Martin Fowler published the first book about refactoring that has as title *Improving the Design of Existing Code* [4]. This book popularised the practice of code refactoring, set its fundamentals, and had a high impact on the world of software development. Martin Fowler defined Refactoring in his book as a sequence of small changes - called refactoring operations - made to the internal structure of the code without altering its external behavior. The goal of these refactoring operations is to improve the code readability and reusability as well as reduce its complexity and maintenance costs in the long run. Since then, a lot has changed in the software development world, but one thing has remained the same: The need for Refactoring.

Nearly 30 years later, Refactoring has become a crucial part of software development practice, especially with the ever-changing landscape of IT and user requirements. It is a core element of agile methodologies, and most professional IDEs include refactoring tools. Recent studies show that restructuring software systems may reduce developers' time by over 60% [5]. Others demonstrate how Refactoring can help detect, fix, and reduce software bugs [6]. Companies are becoming more and more aware of the importance of Refactoring, and they encourage their developers to continuously refactor their code to set a clean foundation for future updates.

It might be difficult for a developer to be justified to spend time on improving a piece of code to have the same functionality. However, it can be seen as an investment for future developments. Specifically, Refactoring is a crucial task on software with longer lifespans with multiple developers need to read and understand the codes. Refactoring can improve both the quality of software and the productivity of its developers. Increasing the quality of the software is due to decreasing its complexity at design and source code level caused by refactoring, which is proved by many studies [7], [8]. The long-term effect of Refactoring is improving developers' productivity by increasing two crucial factors, understandability and maintainability of the codes, especially when a new developer joins an existing project. It is shown that Refactoring can help to detect, fix, and reduce software bugs and leading to software projects which are less likely to expose bug in development process [6]. Another study claims that there are some specific kinds of refactoring methods that are very probable to induce bug fixes [9].

- *Chaima Abid, Vahid Alizadeh, Marouane Kessentini, and Thiago do Nascimento Ferreira are with the department of Computer and Information Science, University of Michigan, Dearborn, MI, USA.*
  *Danny Dig is with the Computer Science department, University of Colorado, Boulder, CO, USA.*
  *E-mail: marouane@umich.edu*

### 1.1 Problem Description and Motivation

Refactoring is among the fastest-growing software engineering research areas, if not the fastest. Figure 1 shows the distribution of publications related to refactoring across the globe. Figure 2 reflects the number of publications in the top 10 most active countries in the field of Refactoring. The United States tops the list of countries with a total of 714 publications followed by Germany and Canada with a total of 317 and 248 publications, respectively. During the past 4 years, the number of published refactoring studies has increased with an average of 37% in all top 10 countries. This demonstrates a noticeable increase in interest/need in Refactoring.

Over 5584 authors from all over the world contributed to the field of Refactoring. We highlight the most active authors in Figure 3 and 4, based on both the number of publications and citations in the area. Many scholars started research in the refactoring filed prior to 2000. Others are relatively new to the field and started their contributions after year 2010. All top 10 authors in the field have a constantly increasing number of publications over the past 20 years. Marouane Kessentini heads the list with a total of 43 publications (51% of them were published during the past five years) followed by Steve Counsell and Danny Dig with a total of 39 and 36 publications, respectively. Marouane kessentini published an average of more than 4 articles per year while all other authors published an average between 1.5 and 2.75 publications per year. Figure 5 is a histogram showing how many publications were issued each year starting from 1990. The number of published journal articles, conference papers, and books has increased dramatically during the last decade, reaching a pick of 265 publications in 2016. During just the last four years (2016-2019), over 1026 papers were published in the field, with an average of 256 papers each year.

Recently, several researchers and practitioners have adopted the use of refactoring operations at higher degrees of abstraction than source code level (e.g., databases, Unified Modeling Language (UML) models, Object Constraint Language (OCL) rules, etc.). As a result, they often had to redefine the principles and guidelines of refactoring according to the requirements and specifications of their domains. For instance, in User Interface Refactoring, developers make changes to the UI to retain its semantics and consistency for all users. These refactorings include, but not limited to, *Align entry field, Apply common button size, Apply font, Indicate format,* and *Increase color contrast*. In Database Refactoring, developers improve the database schema by applying changes such as *Rename column, Split table, Move method, Replace LOB with table,* and *Introduce column constraint*. Henceforth, the refactoring operations are called restructuring operations when applied to artifacts other than the ones related to object-oriented programming. Although the different refactoring communities (e.g., software maintenance and evolution, model-driven engineering, formal methods, search-based software engineering, etc.) are interdependent in many ways, they remain disconnected, which may create inconsistencies. For example, when model-level Refactoring does not match the code-level practice, it can lead to incoherence and technical issues during development. The detachment is visible not only between different refactoring domains but also between practitioners and researchers. The distance between them primarily originates from the lack of insights into both worlds' recent findings and needs. For instance, developers tend to use the refactoring features provided by IDEs due to their accessibility and popularity. Most of the time, they are uninformed of the benefits that can be derived from adopting state-of-the-art advances in academia. All these challenges call for a need to identify, critically appraise, and summarize the existing work published across the different domains. Existing systematic literature reviews examine findings in very specific refactoring areas such as identifying the impact of refactoring on quality metrics [10] or code smells [11]. To the best of our knowledge, no work collects and synthesizes existing research, tools, and recent advances made in the refactoring community. This paper is the most comprehensive synthesis of theories and principles of refactoring intended to help researchers and practitioners make quick advances and avoid reinventing or re-implementing research infrastructure from scratch, wasting time and resources. We also build a refactoring infrastructure that will connect researchers with practitioners in industry and provide a bridge between different refactoring communities in order to advance the field of refactoring research.

### 1.2 Contributions

The Refactoring area is growing very rapidly, and many advances, challenges, and trends have lately emerged. The primary purpose of this study is to implement a systematic literature review (SLR) for the field of refactoring as a whole. This SLR follows a defined protocol to increase the study's validity and rationality so that the output can be high in quality and evidence-based. We used various electronic databases and a large number of articles to comprise all the possible candidate studies and cover more works than existing SLRs.

This SLR contributes to the existing literature in the following ways:

- We identify a set of 3183 studies related to refactoring published until May 2020, fulfilling the quality assessment criteria. These studies can be used by the research and industry communities as a reliable basis and help them conduct further research on Refactoring.
- We present a comprehensive qualitative and quantitative synthesis reflecting the state-of-the-art in refactoring with data extracted from those 3183 high-rigor studies. Our synthesis covers the following themes: artifacts, refactoring tools, different approaches, and performance evaluation in refactoring research.
- We provide guidelines and recommendations based on our findings to support further research in the area.
- We implement a platform that includes the following components: (1) A searchable repository of refactoring publications based on our proposed taxonomy; (2) A searchable repository of authors who contributed to the refactoring community; (3) Analysis and visualization of the refactoring trends and techniques based on the collected papers. The proposed infrastructure will allow researchers and practitioners to easily report refactoring publications and upload information about active authors in the field of Refactoring. It will also bridge the different communities to advance the field of refactoring research and provide opportunities to educate the next refactoring generation.

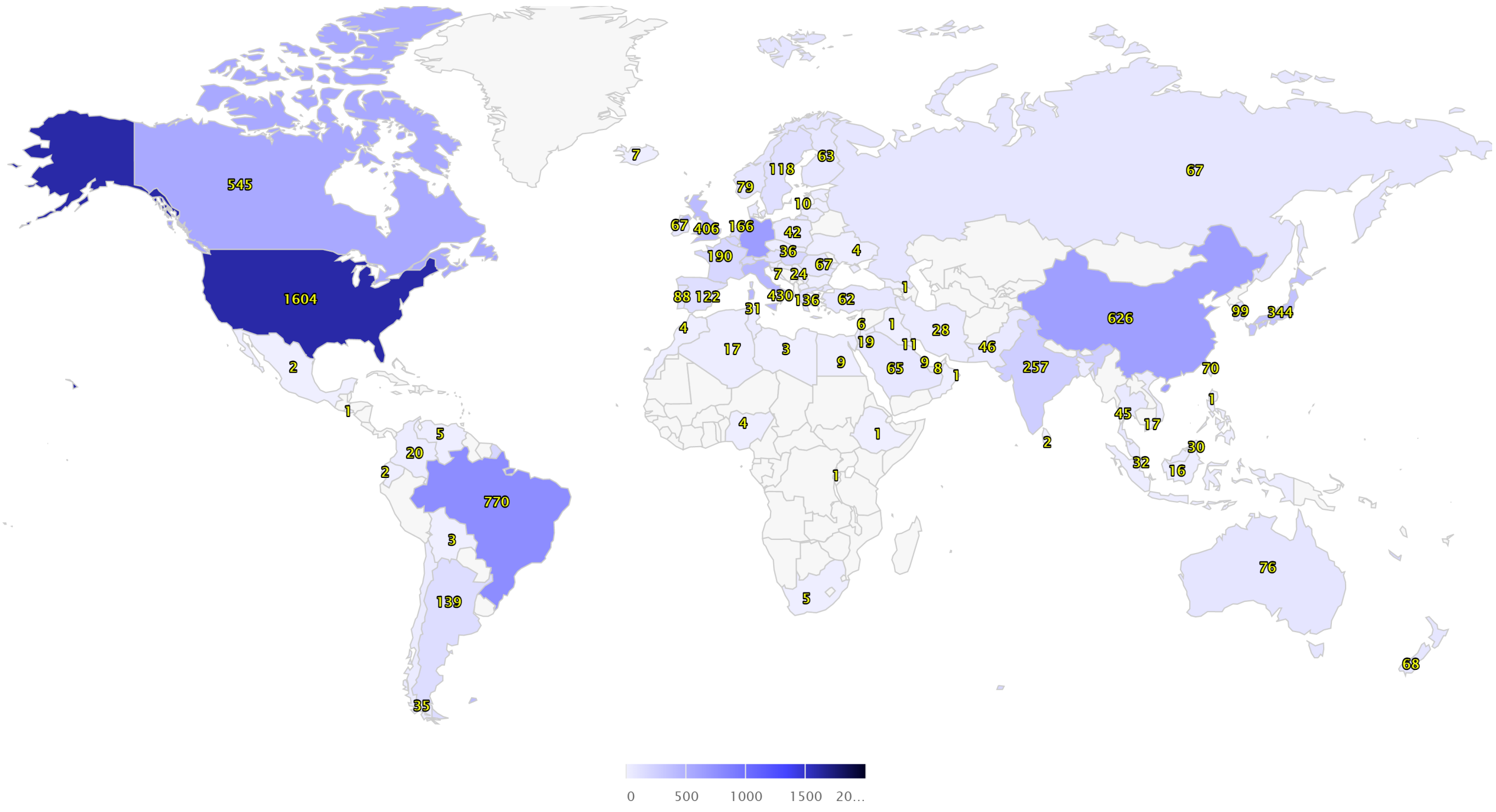

Fig. 1. Distribution of refactoring publications around the world.

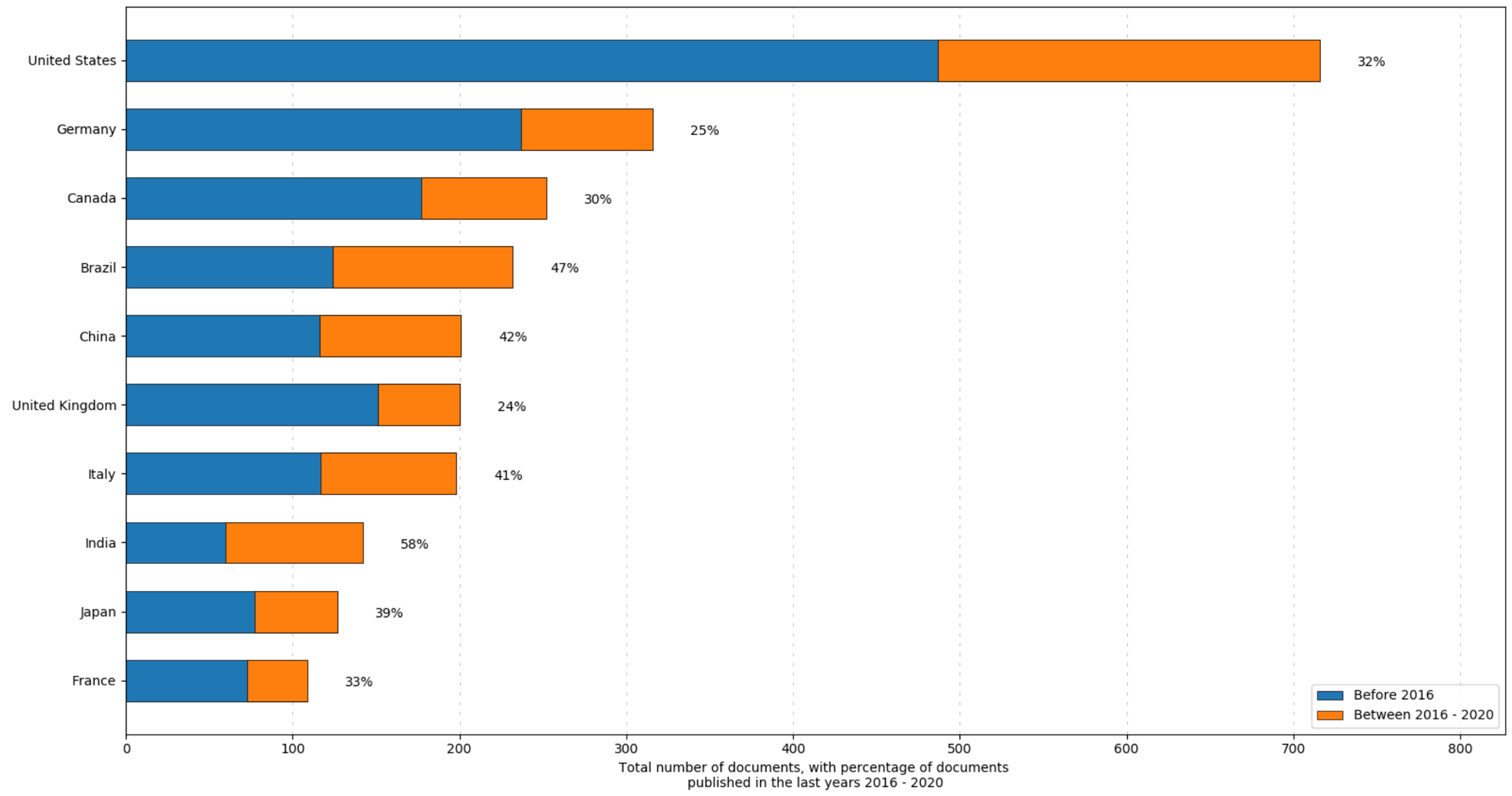

Fig. 2. Number of publications in the top 10 most active countries in the refactoring field

### 1.3 Related Surveys

Mens et al. [12] provided an overview of existing research in the field of software refactoring. They compared and discussed different approaches based on different criteria such as refactoring activities, techniques and formalisms, types of software artifacts that are being refactored, and the effect of refactoring on the software process. Elish et al. [13] proposed a classification of refactoring methods based on their measurable effect on software quality attributes. The investigated software quality attributes are adaptability, completeness, maintainability, understandability, reusabil-ity, and testability. Du Bois et al. [14] provided an overview of the field of software restructuring and Refactoring. They summarized Refactoring's current applications and tool support and discussed the techniques used to implement refactorings, refactoring scalability, dependencies between refactorings, and application of refactorings at higher levels of abstraction. Mens et al. [15] identified emerging trends in refactoring research (e.g., refactoring activities, techniques, tools, processes, etc.), and enumerates a list of open questions, from a practical and theoretical point of views. Misbhauddin et al. [16] provide a systematic overview of existing research in the field of model Refactoring. Al Dallal et al. [17] presented a systematic literature review of existing studies, published through the end of 2013, identifying opportunities for code refactoring activities. In another of their work [10], they presented a systematic literature review that summarizes the impact of refactoring on several internal and external quality attributes. Singh et al. [11] published a systematic literature review of refactoring concerning code smells. However, the review of Refactoring is done in a general manner, and the identification of code smells and anti-patterns is performed in-depth. Abebe et al. [18] conducted a study to reveal the trends, opportunities, and challenges of software refactor researches using a systematic literature review. Baqais et al. [19] performed a systematic literature review of papers that suggest, propose, or implement an automated refactoring process.

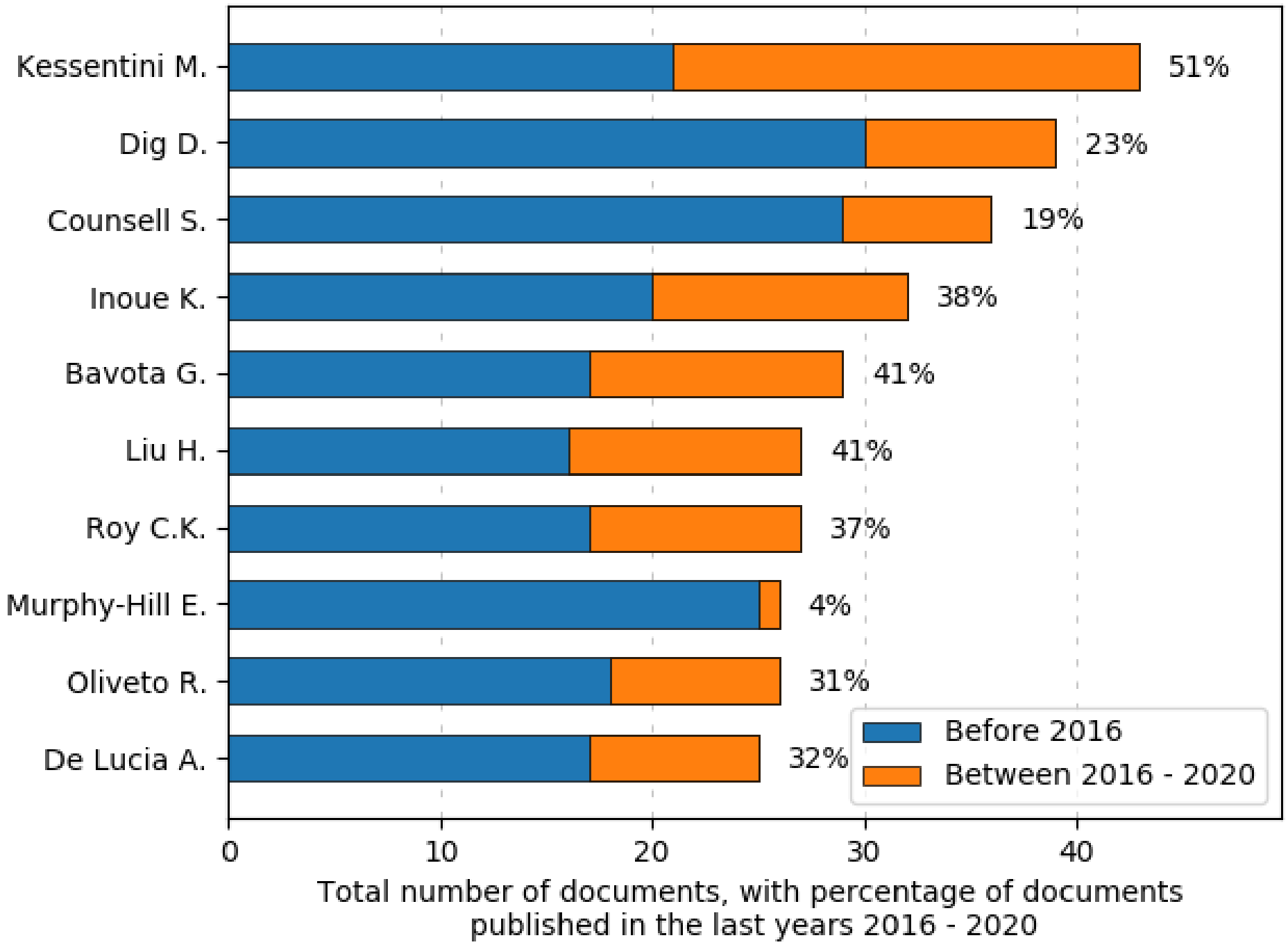

Fig. 3. Top 10 Authors with the highest number of publications and citations in the field of refactoring

The different studies mentioned above are mainly about identifying the studies related to very specific or specialized topics. In this paper, we are trying to be as comprehensive as possible by collecting, categorizing, and summarizing all the papers related to refactoring in general that conform to our quality standards.

### 1.4 Organization

The rest of the paper is organized as follows: First, Section 2 outlines the research method and the underlying protocol for the systematic literature review. Section 3 describes the proposed refactoring infrastructure. The results of this systematic review are reported in Sections 4. Finally, Section 5 presents the conclusions.

## 2 Research Methodology

Our literature review follows the guidelines established by Kitchenham and Charters [20], which decompose a systematic literature review in software engineering into three stages: planning, conducting, and reporting the review. We have also taken inspiration from recent systematic literature reviews in the fields of empirical software engineering [10] and search-based software engineering [21]. All the steps of our research are well documented, and all the related data are available online for further validation and exploration []. This section details the performed research steps and the protocol of the literature review. First, section 2.1 describes the research questions underlying our survey. Second, section 2.2 details the literature search step. Next, section 2.3 highlights the inclusion and exclusion criteria. The data preprocessing step and our proposed taxonomy are described in sections 2.4 and 2.5, respectively. The quality assessment criteria are defined in section 2.6. Finally, Section 2.7 discusses threats to the validity of our study.

### 2.1 Research Questions

The following research questions have been derived based on the objectives described in the introduction, which form the basis for the literature review:

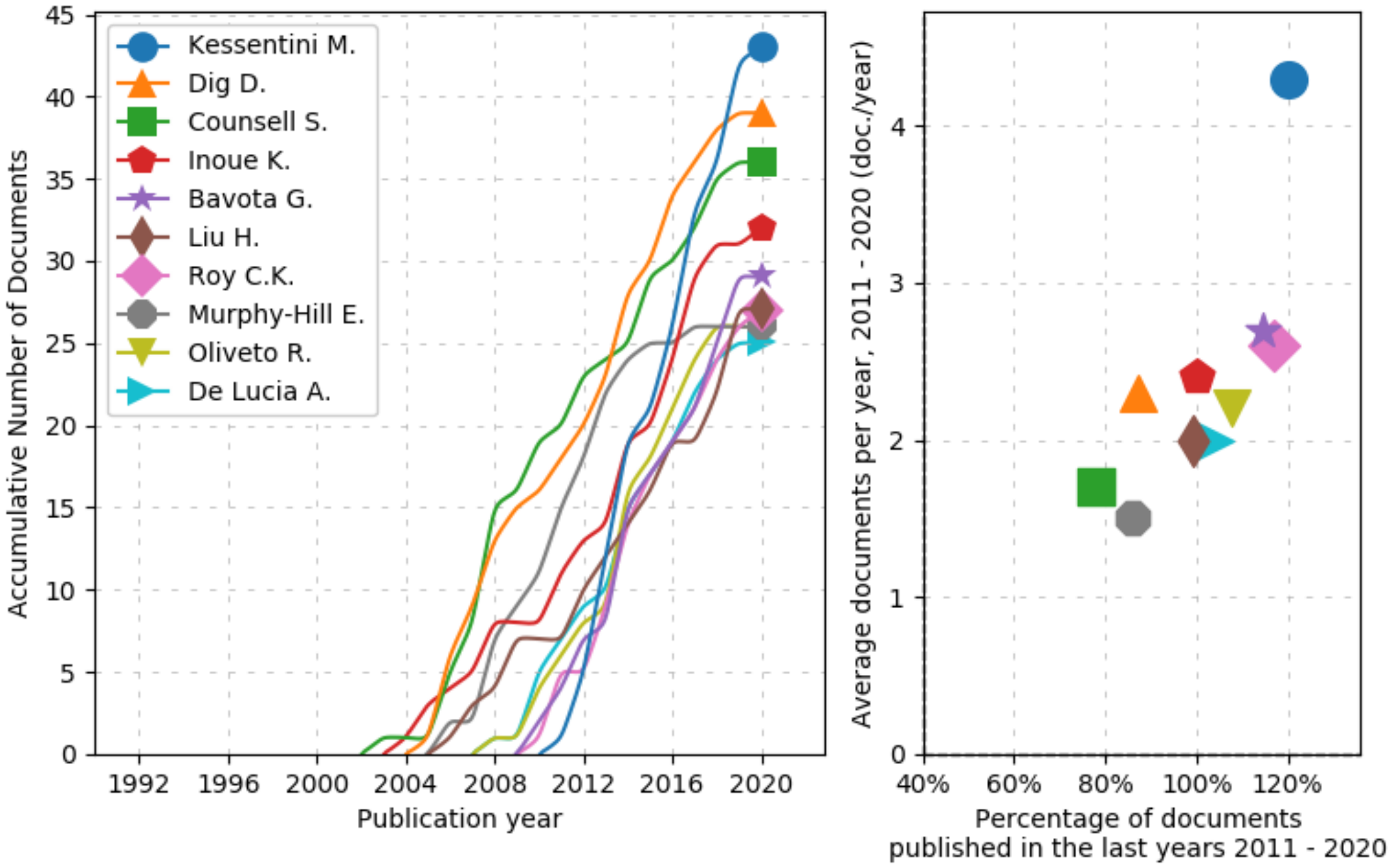

Fig. 4. Evolution of the Top 10 Authors during the past 10 years

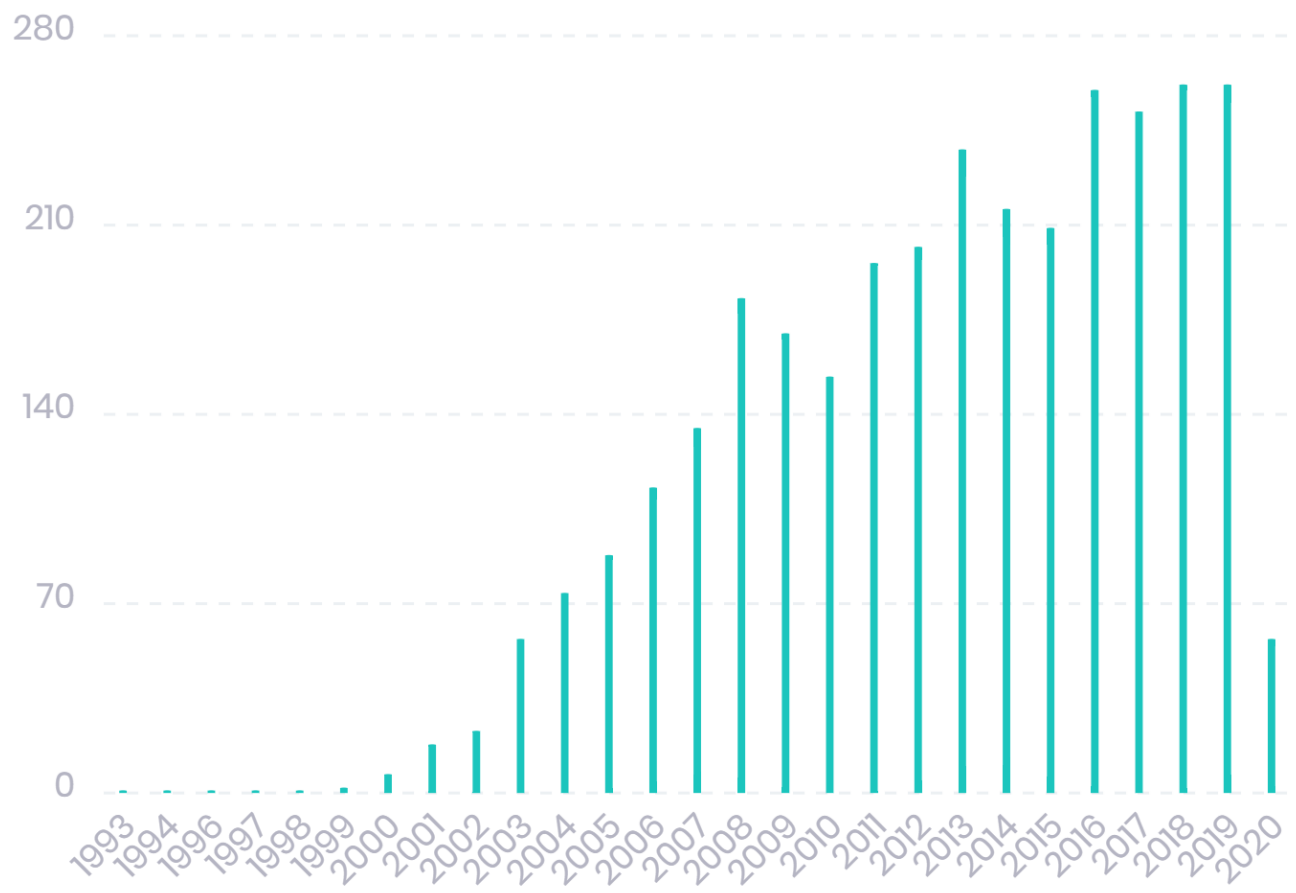

Fig. 5. Trend of publications in the field of refactoring during the last three decades.

- RQ1: What is the refactoring life-cycle?
- RQ2: What are the types of artifacts that are being refactored at each step of the refactoring life-cycle?
- RQ3: Why do software practitioners and researchers perform refactoring?
- RQ4: What are the different approaches used by software practitioners and researchers to perform refactoring?
- RQ5: What types of datasets are used by software practitioners and researchers to validate the refactoring?

### 2.2 Literature Search Strategy

All the papers have been queried from a wide range of scientific literature sources to make our search as comprehensive as possible:

- **Digital libraries:** ACM Library, IEEE Xplore, ScienceDirect, SpringerLink.
- **Citation databases:** Web of Science (formerly ISI Web of Knowledge), Scopus.
- **Citation search engines:** DBLP, Google Scholar.

We first defined a list of terms covering the variety of both application domains and refactoring techniques. For that, we checked the title, keywords, and abstract of the relevant papers that were already known to us. Synonyms and keywords were derived from this list. These keywords were combined using logical operators ANDs and ORs to create search terms. Before starting collecting the primary studies (PS), we tested the search terms' effectiveness on all the data sources. Then, we refined the queries to avoid getting irrelevant papers. The string adjustments were agreed on by all authors. The final list of search strings are shown in Table 1. These search strings were modified to suit the specific requirements of different electronic databases. We conducted our search on May 31st, 2020, and identified studies published up until that date. The search was done first by the corresponding author and then verified by the rest of the authors. In our systematic review, we followed a multi-stage model to minimize the probability of missing relevant publications as much as possible. The different stages are shown in figure 6 along with the total returned publications at each stage. The first stage consists of executing the search queries on the databases mentioned above; a total of 6158 references were found. Then, we removed the duplicates, which reduced the list of candidate papers to 3882. Then, we performed a manual examination of titles and abstracts to discard irrelevant publications based on the inclusion and exclusion criteria. We also looked at the body of the paper whenever necessary. This decreased the list of candidate papers to 3161 publications. Next, we used the resulting set as input for the snowballing process, recommended by Wohlin [22], to identify additional studies. We consulted web profiles of relevant authors and their networks. We also checked cross-references until no further papers were detected. As a result, 17 new references were added. After that, we contacted the corresponding authors of the identified publications to inquire about any missing relevant studies. This led to adding 5 studies.

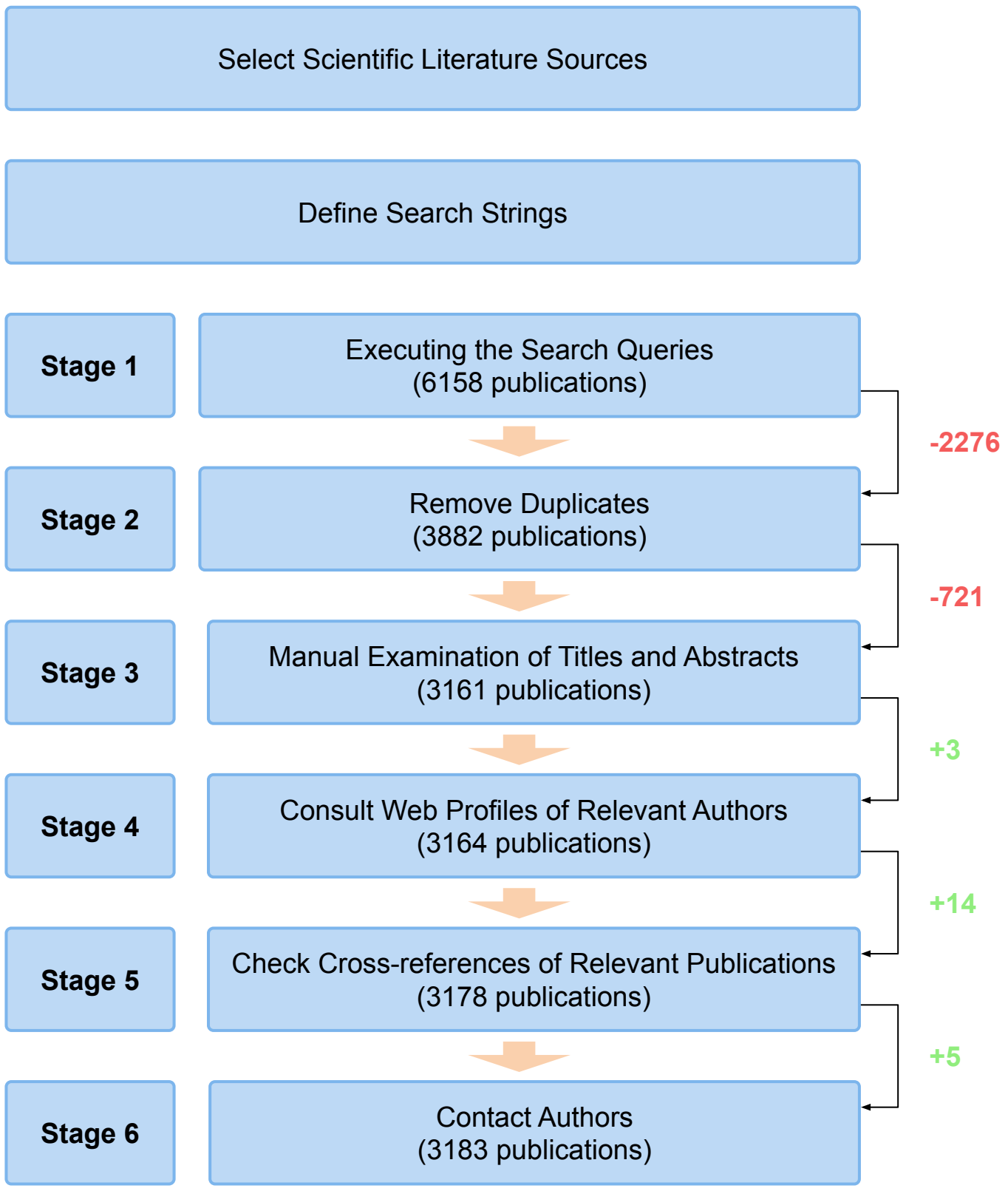

Fig. 6. SLR steps

## 2.3 Inclusion and Exclusion Criteria

To filter out the irrelevant articles among those selected in Stage 2 and determine the Primary studies, we considered the following inclusion and exclusion criteria.

### *2.3.1 Inclusion criteria*

All of the following criteria must be satisfied in the selected primary studies:

1) The article must have been published in a peer reviewed journal or conference proceeding between the years 1990 and 2020. The main reason for imposing a constraint over the start year is because the first known use of the term “refactoring” in the published literature was in a September, 1990 article by William Opdyke and Ralph Johnson [1]. We included papers up till May 31st 2020.
2) The article must be related to computer science and engineering and propose techniques, methods and tools for refactoring.
3) The paper must be written in English.
4) In case a conference paper has a journal extension, we would include both the conference and journal publications.
5) The paper must pass the quality assessment criteria that are elaborated in Section 2.6.

### *2.3.2 Exclusion criteria*

Papers satisfying any of the exclusion criteria were discarded, as follows:

1) Studies that are not related to the computer science field.
2) Studies that investigated the impact of general maintenance on code quality. In this case, the maintenance tasks were potentially performed due to several reasons and not limited to refactoring, and therefore, we cannot judge whether the impact was due to refactoring or to other maintenance tasks such as corrective or adaptive maintenance.
3) Grey Literature

## 2.4 Data Preprocessing

A pre-processing technique was applied to improve reliability and precision, as detailed in the following sub sections.

### *2.4.1 Simplifying Author’s name*

In general, scientific and bibliographic databases such as *Web of Science* (WoS) and *Scopus* have the following inconsistencies in authors names:

- Most journals abbreviate the author’s first name to an initial and a dot.
- Most journals use the author name’s special accents.
- WoS uses a comma between the author’s last name and first name initial, but Scopus does not.

These name-related inconsistencies mean that scientometrics scripts cannot find all of the similar author’s names. For that reason, ScientoPy script applies the following steps to simplify author’s name fields:

- Remove dots and coma from author’s name.
- Remove special accents from author’s name

### *2.4.2 Fixing inconsistent country names*

Some authors use different naming to refer to the same country (such as USA and United States). For that reason, some country names were replaced based on Table 3.

## 2.5 Study Classification

According to the research questions listed in Section 2.1, we classified the PSs into five dimensions: (1) refactoring lifecycle (related to RQ1), (2) artifacts affected by refactoring (related to RQ2), (3) refactoring objectives (related to RQ3), (4) refactoring techniques (related to RQ4) and (5) refactoring evaluation (related to RQ5). The determination of the attributes of each dimension was performed incrementally. That is, for each dimension, we started with an empty set of attributes. The authors of this study screened the full texts of the PSs one by one, analyzed each reported study based on the considered dimension, and determined

TABLE 1
final list of search strings

| search strings |
|---|
| (software OR system OR code OR service OR diagram OR database OR architecture OR Model OR GUI OR user interface OR UI OR design OR artifact OR developer OR computer OR programming OR object-oriented OR implement OR mobile app OR cloud OR document ) AND (refactor OR refactoring) |

TABLE 2
PS quality assessment questions [17]

| | Question |
|---|---|
| Design | Are the applied identification techniques for refactoring opportunities clearly described? |
| | Are the refactoring activities considered clearly stated and defined? |
| | Was the sample size justified? |
| | Are the evaluation measures fully defined? |
| Conduct | Are the data collection methods adequately described? |
| Analysis | Are the results of applying the identification techniques evaluated? |
| | Are the data sets adequately described? (size, programming languages, source) |
| | Are the study participants or observational units adequately described? |
| | Are the statistical methods described? |
| | Are the statistical methods justified? |
| | Is the purpose of the analysis clear? |
| | Are the scoring systems (performance evaluation) described? |
| Conclusion | Are all study questions answered? |
| | Are negative findings presented? |
| | Are the results compared with previous reports? |
| | Do the results add to the literature? |
| | Are validity threats discussed? |

TABLE 3
List of countries and their replacements

| Country | Replacement |
|---|---|
| Republic of China | China |
| USA | United States |
| England, Scotland and Wales | England |
| U Arab Emirates | United Arab Emirates |
| Russia | Russian Federation |
| Viet Nam | Vietnam |
| Trinid & Tobago | Trinidad and Tobago |

the attributes of that dimension as considered by each PS. Table 4 outlines the keywords extracted for each category. It should be pointed out that, most of the time, we remove all of the affixes (i.e., suffixes, prefixes, etc.) attached to a word in order to keep its lexical base, also known as root or stem or its dictionary form or lemma. For instance, the word *document* allows us to detect the words *documentation* and *documenting*. Also, we did not include bi-grams and tri-grams that can be detected using one uni-gram. For example, *Class Diagram, Object Diagram, Sequence Diagram*, and *Use Case Diagram* can all be detected using the word *Diagram* alone.

The screening of the PSs resulted in determining six stages for the refactoring life-cycle (e.g., detection, prioritization, recommendation, testing, documentation, and prediction). We also classified the papers according to the level of automation of the proposed technique (e.g., automatic, manual, semi-automatic). The results are described in section 4.1. For the second dimension, we identified five artifacts on which the impact of refactoring is studied by at least one of the PSs. These artifacts are code, architecture, model, GUI, and database. The classification of PSs based on these artifacts is discussed in detail in Section 4.2. We subdivided the third dimension into five categories (e.g., External quality, internal quality, performance, migration, and security) to reflect the refactoring objective and six categories (e.g., Object-oriented design, Aspect-oriented design, Model-driven engineering, Documentation, Mobile development, and Cloud computing) to describe the refactoring paradigms. The classification of PSs based on these categories is discussed in detail in Section 4.3. We divided the fourth dimension into four categories (e.g., data mining, search-based algorithms, formal methods, and fuzzy logic) to reveal the refactoring techniques adopted in the studies and into twelve categories (e.g., Java, C, C#, Python, Cobol, PHP, Scala, Smalltalk, Ruby, Javascript, MATLAB, and CSS) to show the most common programming languages used in our PSs. The details of this categorization are reported in section 4.4. Finally, for the fifth dimension, we divide the PSs into two categories: open-source and industrial. The open-source category includes studies that validate their approaches using open source systems. In contrast, the industrial category consists of the studies that validate their work on systems of their industrial collaborators. These findings are outlined in Section 4.5.

## 2.6 Study Quality Assessment

To ensure a level of quality of papers, we only included venues that are known for publishing high-quality software engineering research in general with an h-index of at least 10, as has been done by [23] . Each of the papers that were published before 2019 has to be cited at least once. The quality of each primary study was assessed based on a quality checklist defined by Kitchenham and Charters [20]. This step aims to extract the primary studies with information suitable for analysis and answering the defined research questions. The quality checklist, (described in table 2) were defined by Galster et al. [23]. They are developed by considering bias and validity problems that can occur at different stages, including the study design, conduct, analysis, and conclusion. Each question is answered by a "Yes", "Partially", or "No", which correspond to a score of 1, 0.5, or 0, respectively. If a question does not apply to a study, we do not evaluate the study for that question. The quality assessment checklist was independently applied to all 3882 studies by two of the authors. All disagreements on the quality assessment results were discussed, and a consensus was reached eventually. Few cases where agreement could not be reached were sent to the third author for further investigation. 154 studies did not meet the quality assessment criteria.

TABLE 4
List of keywords used to detect the different categories

| Category | Keywords |
|---|---|
| Refactoring life-cycle (RQ1) | |
| Detection | detect, opportunity, smell, antipattern, design defect |
| Prioritization | schedul, sequence, priorit |
| Recommendation | recommend, correction, correcting, fixing, suggest |
| Testing | test, regression testing, test case, unit test |
| Documentation | document |
| Prediction | predict, future release, next release, development history, refactoring history |
| Level of automation (RQ1) | |
| Manual | manual |
| Semi-automatic | semi-automat, semi-manual |
| Automatic | automat |
| Artifact (RQ2) | |
| Code | code, java, object orient, smell, antipattern, anti-pattern, object-orient |
| Model | design, model, UML, diagram, Unified Modeling Language |
| Architecture | architecture, hotspot, hierarchy |
| GUI | gui, user interface, UI |
| Database | relational, schema, database, Structured Query Language, SQL |
| Paradigm (RQ3) | |
| Object-oriented design | object orient, object-orient, oo, java, c, ++, python, C sharp, c#, css, Python, R, PHP, JavaScript, Ruby, Perl, Object Pascal, Objective-C, Dart, Swift, Scala, Kotlin, Common Lisp, MATLAB, Smalltalk |
| Aspect-oriented design | aspect |
| Model-driven engineering | model transform, uml, reverse engineering, diagram, Unified Modeling Language |
| Documentation | document |
| Mobile development | android, mobile, IOS, phone, smartphone, cellphones |
| Could computing | web service, wsdl, restful, cloud, Apache Hadoop, Docker, Middleware, Software-as-a-Service, SaaS, XaaS, Anything-as-a-Service, Platform-as-a-Service, PaaS, Infrastructure-as-a-Service, IaaS, AWS, Amazon EC2, Amazon Simple Storage Service, S3 |
| Refactoring Objectives (RQ3) | |
| Internal Quality | maintainability, cyclomatic, depth of inheritance, coupling, quality, Flexibility, Portability, Re-usability, Readability, Testability, Understandability |
| Performance | performance, parallel, Response Time, Error Rates, Request Rate, availability |
| External quality | analysability, changeability, time behaviour, resource, Correctness, Usability, Efficiency, Reliability, Integrity, Adaptability, Accuracy, Robustness |
| Migration | migrat |
| Security | secure, safety, Attack surface, virus, hack, vulnerability, vulnerable, spam |
| Programming languages (RQ4) | |
| Java | java |
| C | c, c++ |
| C# | c sharp, c# |
| Python | python |
| CSS | css |
| PHP | php |
| Cobol | cobol |
| Scala | scala |
| Javascript | javascript |
| Ruby | ruby |
| Smalltalk | smalltalk |
| MATLAB | matlab |
| Adopted methods (RQ4) | |
| Search-based algorithms | search, search-base, sbse, genetic, fitness, simulated annealing, tabu search, search space, Hill climbing, Multi-objective evolutionary algorithms, multi objective optimization, multi-objective programming, vector optimization, multi-criteria optimization, multi-attribute optimization, Pareto optimization, Evolutionary Multi-objective Optimization, EMO, Single-Objective Optimization, Many-Objective Optimization, multi objective |
| Data mining | artificial intelligence, ai , machine learning, naive bayes, decision tree, SVM, support vector machine, Cluster, Classification, classify, Association, Neural networks, deep learning, random forest, regression, reinforcement learning, learning |
| Formal methods | model check, formal method, B-Method, RAISE, Z notation, SPARK Ada |
| Fuzzy logic | fuzzy |
| Evaluation method (RQ5) | |
| Open source | open source, open-source |
| Industrial | proprietary, industrial, industry, collaborator, collaboration |

### 2.7 Threats to Validity

Several limitations may affect the generalizability and the interpretations of our results. The first is the possibility of paper selection bias. To ensure that the studies were selected in an unbiased manner, we followed the well-defined research protocol and guidelines reported by Kitchenham and Charters [20] instead of proposing nonstandard quality factors. Also, the final decision on the articles with selection disagreements was performed based on consensus meetings. The Primary studies were assessed by one researcher and checked by the other, a technique applied in similar studies [21]. The second threat consists of missing a relevant study. To overcome this threat, we employed several strategies that we mentioned in Section 2.2. Few related studies were detected after performing the automatic search, which indicates that the constructed search strings and the mentioned utilized libraries were comprehensive enough to identify most of the relevant articles. Another critical issue is whether our taxonomy is complete and robust sufficient to analyze and classify the primary studies. To overcome this problem, we used an iterative content analysis method by going through the papers one by one and continuously expand the taxonomy for every new encountered concept. Furthermore, to gather sufficient keywords to detect the different categories, we followed the same iterative process, and we added synonyms based on the authors' expertise in the field of refactoring. Another threat is related to the tagging of the papers according to our taxonomy. To mitigate this problem, we asked 27 graduate students to check the correctness of the classification results by reading the abstract, the title, and keywords. They also check the body of the paper whenever necessary.

## 3 Refactoring Infrastructure

We implemented a large scale platform [24] that collects, manages, and analyzes refactoring related papers to help researchers and practitioners share, report, and discover the latest advancements in software refactoring research. It includes the following components:

1) **A searchable repository of refactoring publications based on our proposed taxonomy**. Figure 9 shows a screenshot of the publications' tab of the refactoring repository website. The papers can be searched by author, title, or year of publication. Each paper has tags that describe its content based on our taxonomy described in section 2.5. The papers can also be filtered using those tags and sorted alphabetically or chronologically according to the title and year of publication, respectively. The user can export the publications' dataset to many formats, including pdf, excel, and CSV. He can also easily report a new publication by entering its link.
2) **A searchable repository of authors who contributed to the refactoring community**. Figure 8 shows a screenshot of the authors' tab of the refactoring repository website. The authors can be searched and sorted alphabetically by name, affiliation, or country. They can also be sorted based on the total number of refactoring publications. The user can also consult the *Google Scholar* and *Scopus* profiles of the authors if available. Finally, the user can easily report a new author by entering their information and their profile. Furthermore, we defined the refactoring h-index, which shows how many papers about refactoring published by the author have been cited proportionately. A refactoring h-index of X means that the author has X papers about refactoring that have been cited at least X times. Authors can also be sorted according to the refactoring h-index and the total number of citations (see figure 11). Besides, we created a co-author network and corresponding visualizations (see figure 12) to get a snapshot view of the breadth and depth of an individual's collaborations in the field of refactoring research. Finally, we generated a histogram (see figure 7) that shows the number of publications issued by the top institutions active in the refactoring research by considering the authors' affiliations.
3) **Analysis and visualization of the refactoring trends and techniques based on the collected papers**. Figure 10 shows a screenshot of the refactoring repository dashboard. It contains histograms and pie charts that show the distribution and percentages of the categories defined in our taxonomy. It also includes maps that reflect the spread of refactoring activity across the world.

The proposed infrastructure will enable researchers to perform a fair comparison between their new refactoring approaches and state-of-the-art tools; enable researchers to use refactoring data of large software systems; facilitate interactions between researchers from currently disconnected domains/communities of refactoring (model-driven engineering, service computing, parallelism and performance optimization, software quality, testing, etc.); enable practitioners and researchers to quickly identify relevant existing research papers and tools for their problems based on the proposed taxonomy and classification; create benchmarks against which various refactoring approaches can be evaluated; enable effective interactions between practitioners and refactoring researchers to identify relevant problems faced by the software industry.

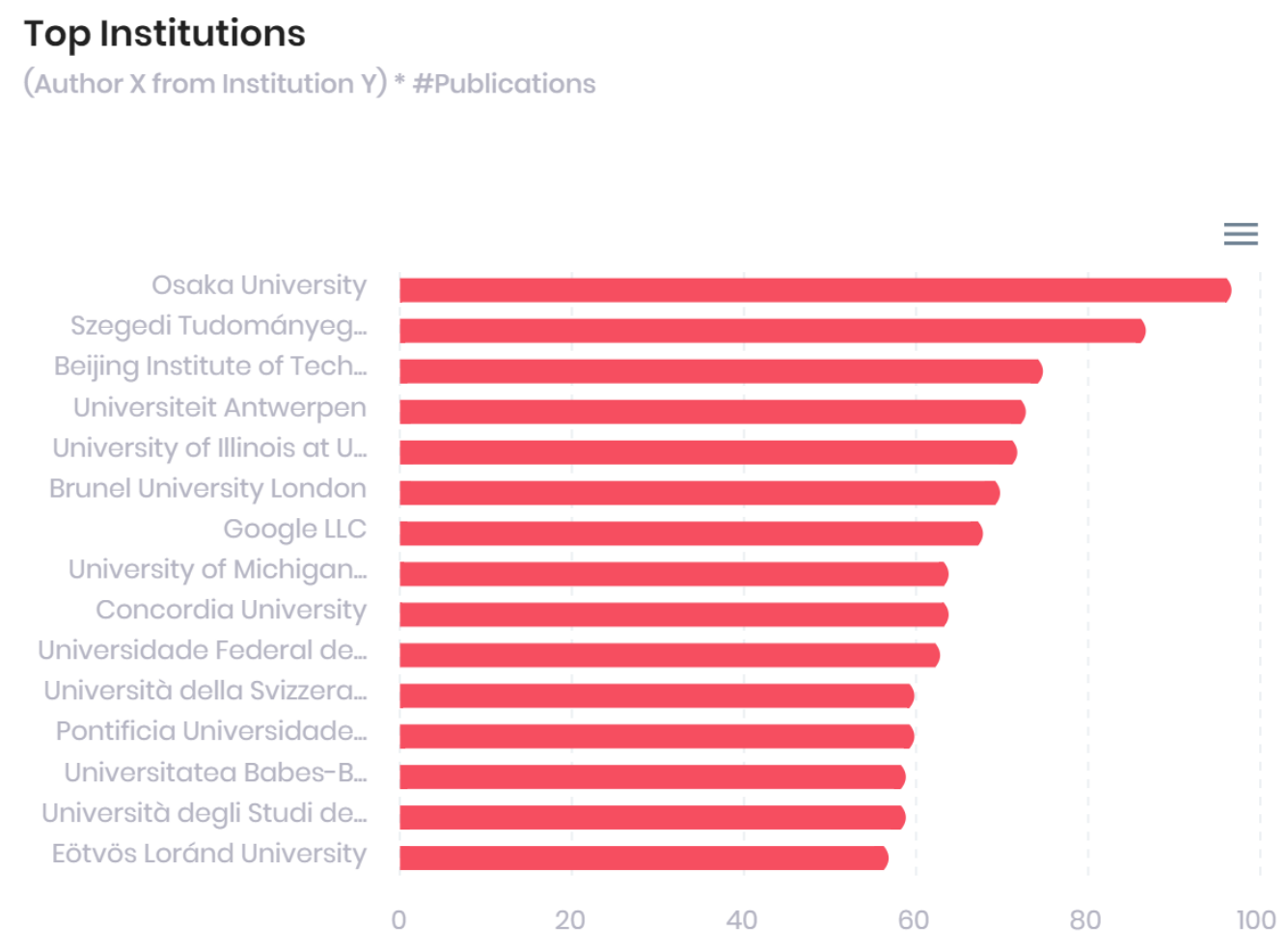

Fig. 7. Top institutions active in the refactoring field

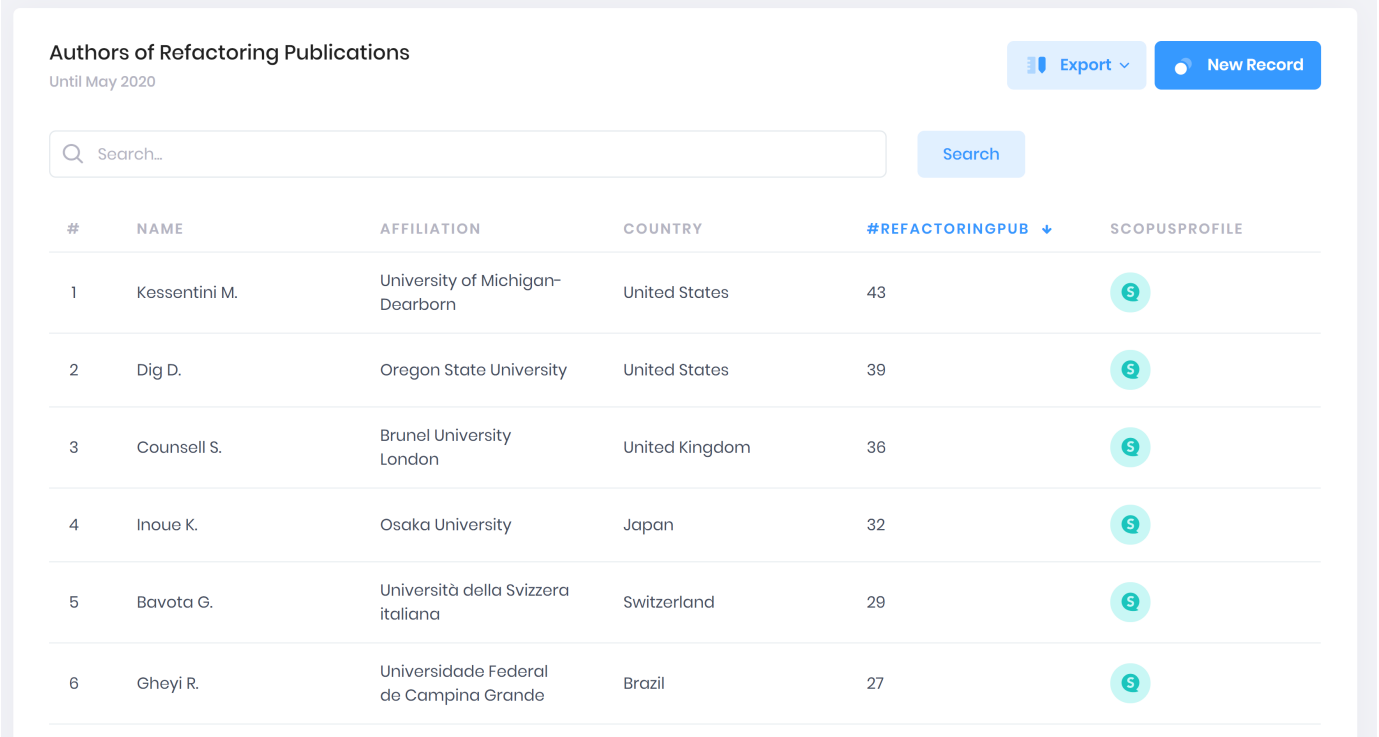

Fig. 8. A screenshot of the authors tab of the refactoring repository Website

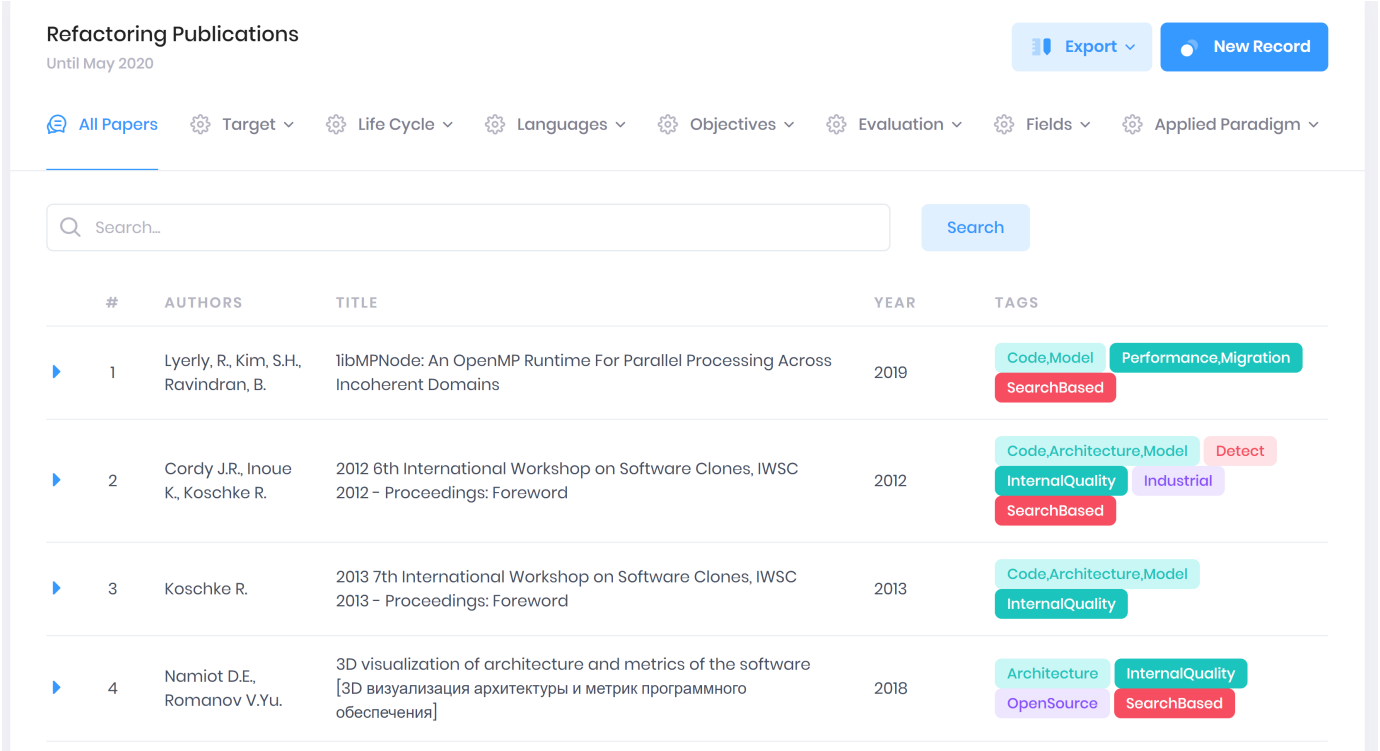

Fig. 9. A screenshot of the publications tab of the refactoring repository Website

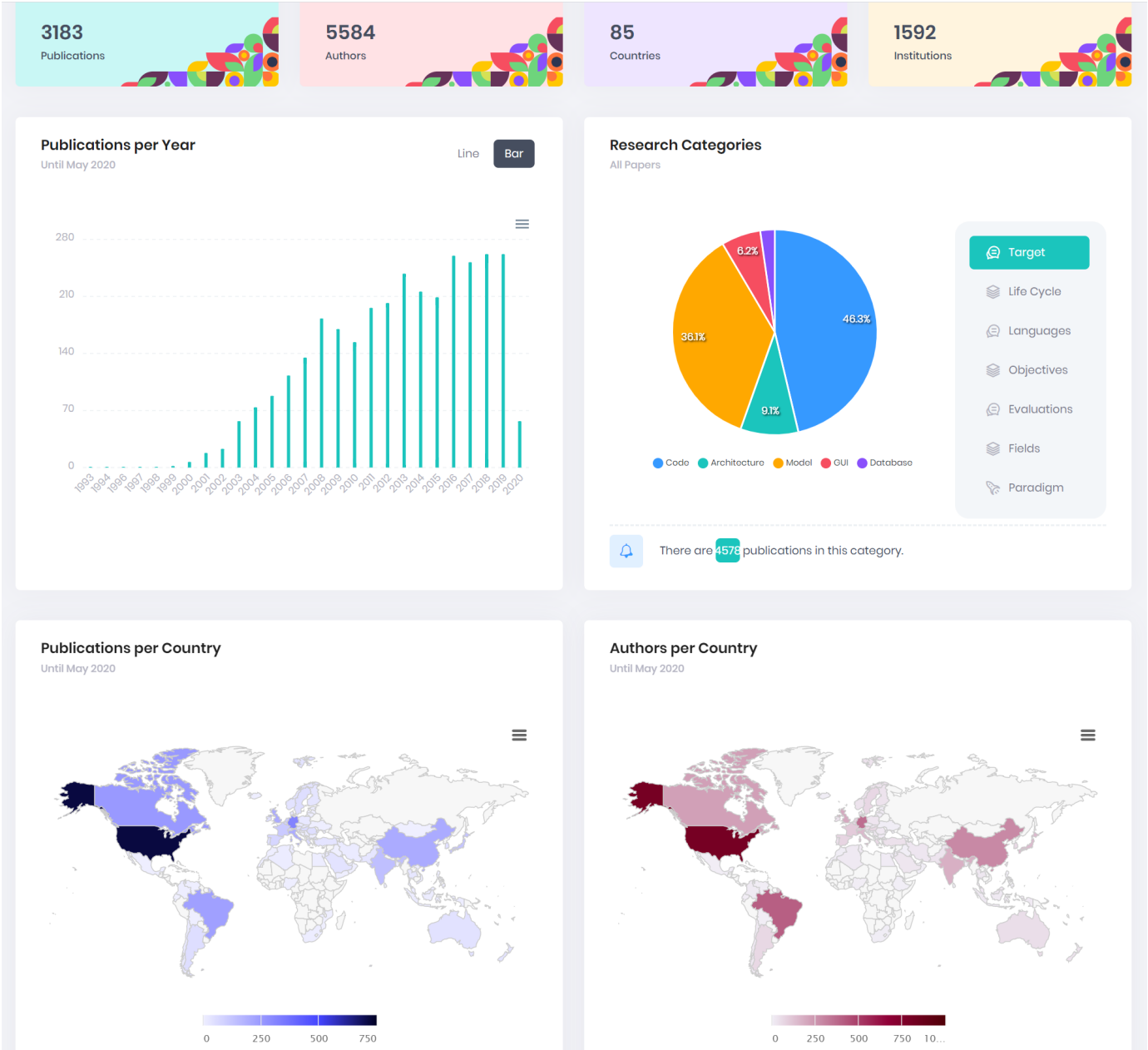

Fig. 10. A screenshot of the Dashboard of the refactoring repository website

## 4 Results

In this section, we aim to answer the research questions. To provide an overview of the current state of the art in refactoring and guide the reader to a specific set of approaches, tools, and recent advances that are of interest, we classified the 3183 reviewed papers based on the taxonomy described in Section 2.5. Table 5 contains representative references for the categories created for each RQ. We only provided 10 references per category because we cannot possibly report in this paper the categorization of all the studies since we are dealing with a total of 3183 papers. The results of the classification of all the papers are provided in our website [24]. For some taxonomy categories, papers may have multiple values and thus be listed several times. As a result, percentages in the tables may sum up to more than 100 percent. Also, not all the papers were classified in all dimensions. Consequently, percentages in one dimension may not sum up to 100 percent. The rest of this section presents the observations and insights that can be derived from the visualization of the categories.

### 4.1 Refactoring life-cycle

Going through the primary studies, we have been able to establish a refactoring life-cycle that is composed of six stages:

- **Refactoring detection**: Identifying refactoring opportunities is an important stage that precedes the actual refactoring process. It can be done by manually inspecting and analyzing an artifact of a system to identify refactoring opportunities. However, this technique is time-consuming and costly. Researchers in this area typically propose fully or semi-automated techniques to identify refactoring opportunities. These techniques may be applicable to different artifacts and should be evaluated empirically.
- **Refactoring prioritization**: The number of refactoring opportunities usually exceeds the amount of problems that the developer can deal with, particularly when the effort available for performing refactorings is limited. Moreover, not all refactoring opportunities are equally relevant to the goals of the system or its health. In this stage, the refactorings operations are prioritized using different criteria (e.g., maximizing the refactoring of classes with a large number of anti-patterns or with the previous history of bugs, etc.) according to the needs of developers.

TABLE 5
Representative references for all categories

| Category | Percentage | Papers |
|---|---|---|
| Refactoring life-cycle (RQ1) | | |
| Detection | 28.65% | [S1], [S2], [S3], [S4], [S5], [S6], [S7], [S8], [S9], [S10] |
| Prioritization | 9.43% | [S11], [S12], [S13], [S14], [S15], [S16], [S17], [S18], [S19], [S20] |
| Recommendation | 16.18% | [S3], [S11], [S12], [S21], [S22], [S23], [S24], [S25], [S26], [S27] |
| Testing | 18.44% | [S4], [S6], [S7], [S8], [S13], [S28], [S29], [S30], [S31], [S32] |
| Documention | 5.22% | [S33], [S34], [S35], [S36], [S37], [S38], [S39], [S40], [S41], [S42], [S43] |
| Prediction | 4.818% | [S44], [S45], [S46], [S47], [S48], [S49], [S50], [S51], [S52], [S53] |
| Level of automation (RQ1) | | |
| Automatic | 30.95% | [S54], [S55], [S56], [S57], [S58], [S59], [S60], [S61], [S62], [S63] |
| Semi-automatic | 1.95% | [S64], [S65], [S66], [S67], [S68], [S69], [S70], [S71], [S72], [S73], [S74], [S75] |
| Manual | 8.67% | [S69], [S76], [S77], [S78], [S79], [S80], [S81], [S82], [S83], [S84] |
| Artifact (RQ2) | | |
| Code | 72.89% | [S1], [S2], [S3], [S11], [S65], [S85], [S86], [S87], [S88], [S89] |
| Model | 59.25% | [S1], [S3], [S28], [S29], [S65], [S87], [S89], [S90], [S91], [S92] |
| Architecture | 17.25% | [S28], [S91], [S93], [S94], [S95], [S96], [S97], [S98], [S99], [S100] |
| GUI | 2.58% | [S6], [S8], [S28], [S87], [S89], [S90], [S101], [S102], [S103], [S104] |
| Database | 4.12% | [S27], [S36], [S65], [S100], [S105], [S106], [S107], [S108], [S109], [S110] |
| Paradigm (RQ3) | | |
| Object-oriented design | 34.09% | [S1], [S8], [S30], [S85], [S87], [S88], [S101], [S111], [S112], [S113] |
| Aspect-oriented | 10.87% | [S88], [S96], [S101], [S102], [S103], [S114], [S115], [S116], [S117], [S118] |
| Model-driven engineering | 7.35% | [S3], [S15], [S32], [S58], [S65], [S119], [S120], [S121], [S122], [S123] |
| Mobile apps development | 3.55% | [S23], [S87], [S87], [S95], [S99], [S112], [S124], [S125], [S126], [S127] |
| Could computing | 4.15% | [S128], [S129], [S130], [S131], [S132], [S133], [S134], [S135], [S136], [S137] |
| Refactoring Objective (RQ3) | | |
| Internal Quality | 41.63% | [S3], [S12], [S21], [S29], [S30], [S89], [S90], [S94], [S138], [S139] |
| Performance | 15.93% | [S10], [S12], [S28], [S86], [S88], [S91], [S92], [S96], [S115], [S119] |
| External quality | 22.68% | [S87], [S91], [S92], [S95], [S102], [S140], [S141], [S142], [S143], [S144] |
| Migration | 3.61% | [S95], [S100], [S113], [S145], [S146], [S147], [S148], [S149], [S150], [S151] |
| Security | 3.11% | [S113], [S152], [S153], [S154], [S155], [S156], [S157], [S158], [S159], [S160] |
| Programming language (RQ4) | | |
| Java | 17.15% | [S1], [S8], [S10], [S30], [S85], [S87], [S88], [S112], [S113], [S140] |
| C | 4.65% | [S59], [S96], [S104], [S105], [S111], [S146], [S161], [S162], [S163], [S164] |
| C# | 0.66% | [S61], [S165], [S166], [S167], [S168], [S169], [S170], [S171], [S172], [S173] |
| Python | 0.53% | [S174], [S175], [S176], [S177], [S178], [S179], [S180], [S181], [S182], [S183] |
| CSS | 0.5% | [S147], [S184], [S185], [S186], [S187], [S188], [S189], [S190], [S191], [S192] |
| PHP | 0.35% | [S169], [S193], [S194], [S195], [S196], [S197], [S198], [S199], [S200], [S201] |
| Cobol | 0.31% | [12], [S202], [S203], [S205], [S206], [S207], [S208], [S209] |
| MATLAB | 0.28% | [S210], [S211], [S212], [S213], [S214], [S215], [S216], [S217] |
| Smalltalk | 0.79% | [25], [S219], [S220], [S221], [S222], [S223], [S224], [S225], [S226], [S227] |
| Ruby | 0.22% | [S169], [S181], [S228], [S229], [S230], [S231] |
| Javascript | 0.72% | [S112], [S232], [S233], [S234], [S235], [S236], [S237], [S238], [S239], [S240], [S241] |
| Scala | 4.02% | [S33], [S55], [S86], [S126], [S242], [S243], [S244], [S245], [S246], [S247] |
| Adopted Method (RQ4) | | |
| Search-based algorithms | 25.76% | [S12], [S248], [S249], [S250], [S251], [S252], [S253], [S254], [S255], [S256] |
| Data mining | 15.49% | [S2], [S82], [S107], [S185], [S257], [S258], [S259], [S260], [S261], [S262] |
| Formal methods | 2.92% | [S42], [S199], [S263], [S264], [S265], [S266], [S267], [S268], [S269] |
| Fuzzy logic | 0.28% | [S257], [S270], [S271], [S272], [S273], [S273], [S274] |
| Evaluation method (RQ5) | | |
| Open source | 16.31% | [S1], [S7], [S12], [S30], [S32], [S88], [S112], [S139], [S248], [S275] |
| Industrial | 10.4% | [S9], [S12], [S16], [S115], [S120], [S147], [S276], [S277], [S278], [S279] |

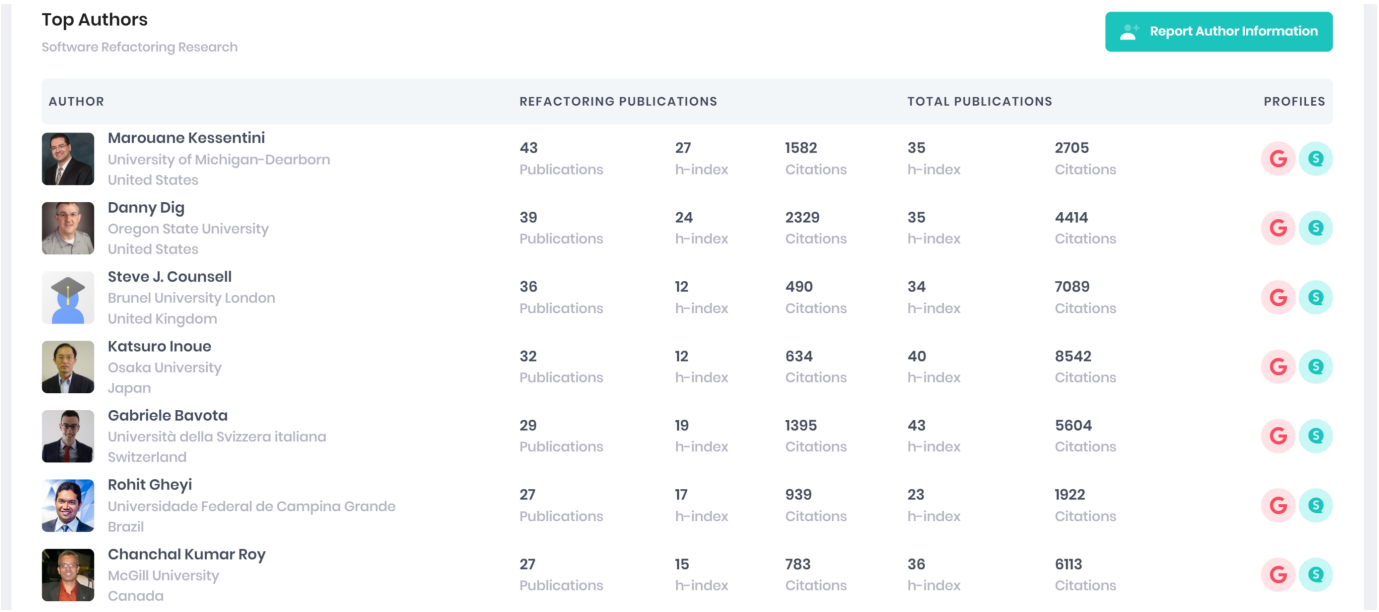

Fig. 11. A screenshot of the refactoring repository dashboard that shows the authors, their h-index and total number of publications and citations

- **Refactoring recommendation**: Several refactoring recommendation tools have been proposed that dynamically adapt and suggest refactorings to developers. The output is sequences of refactorings that developers can apply to improve the quality of systems by fixing, for example, code smells or optimizing security metrics.
- **Refactoring testing**: After choosing the refactorings to be applied, tests need to be done to ensure the correctness of artifacts transformations and avoid future bugs. This is done by checking the satisfaction of the pre-and post-conditions of the refactoring operations and the preservation of the system behavior.
- **Refactoring documentation**: After applying and testing the refactorings, we need to document the refac-

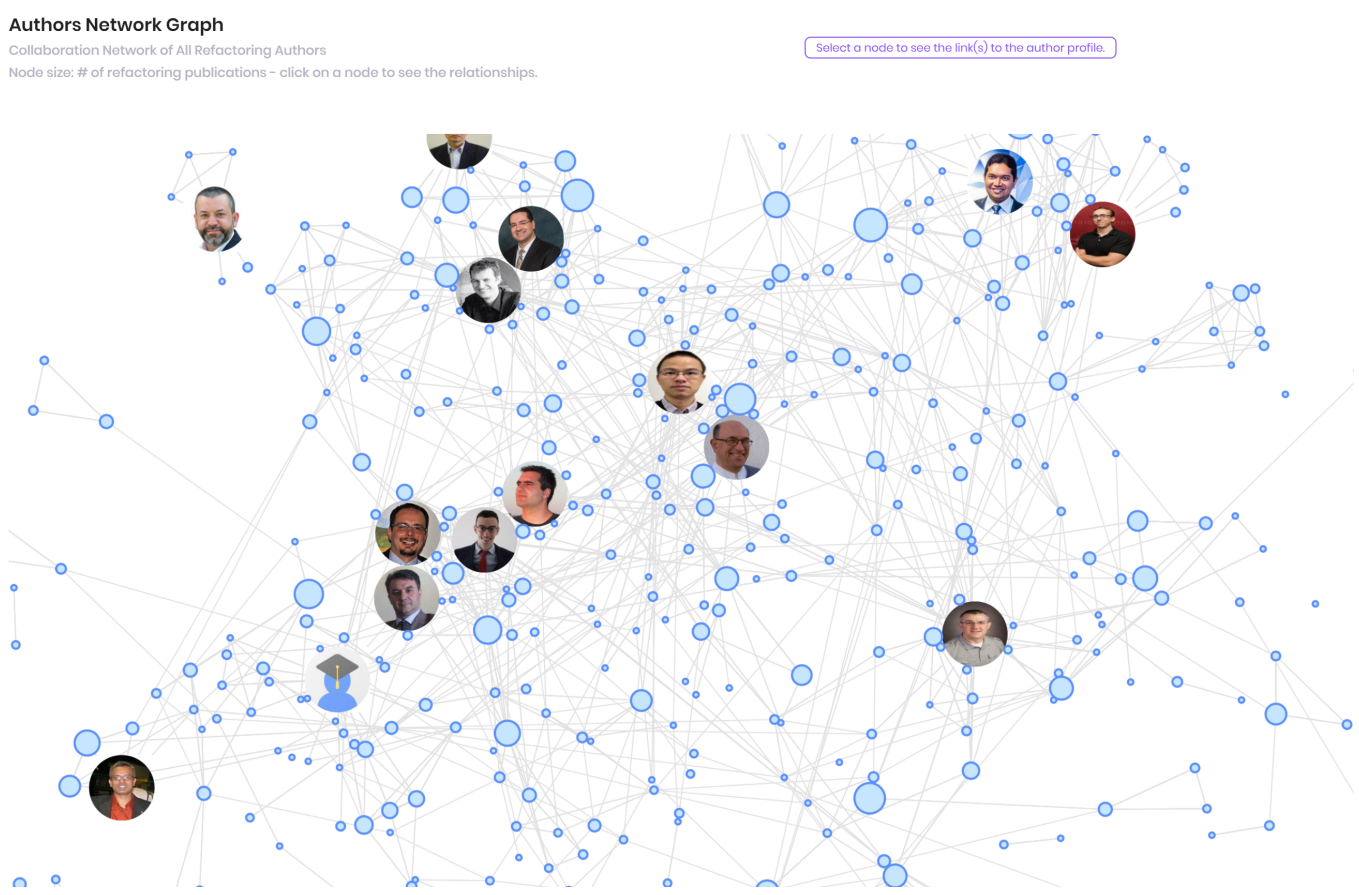

Fig. 12. A screenshot of the authors network graph from the refactoring repository website

torings, their locations, why they have been applied, and the quality improvements.

- **Prediction**: It is interesting for developers to know which locations are likely to demand refactoring in future releases of their software products. This will help them focus on the relevant artifacts that will undergo changes in the future, prepare them for further improvements and extensions of functionality, and optimize the management of limited resources and time. Predicting locations of future refactoring can be done using the development history.

Figure 13 illustrates the percentage of the papers related to each stage of the refactoring life-cycle. 33.08% of the papers deal with testing. Researchers have invested heavily in testing to ensure the reliability of refactoring because changing the structure of code can easily introduce bugs in the program and lead to challenging debugging sessions. A plenty of effort is made towards the automation of the testing process to facilitate the adoption of refactoring [S54], [S55], [S56]. Detecting refactoring opportunities is also a topic of interest to researchers. Several approaches have been proposed to detect refactoring opportunities including but not limited to techniques that depend on quality metrics (e.g., cohesion, coupling, lines of code, etc.), code smells (e.g., feature envy, Blob class, etc.), Clustering (similarities between one method and other methods, distances between the methods and attributes, etc.), Graphs (e.g., represent the dependencies among classes, relations between methods and attributes, etc.), and Dynamic analysis (e.g., analyzing method traces, etc.). Refactoring documentation is an underexplored area of research. Only 5.22% of the collected papers dived into refactoring documentation. Many studies examined the automation of the different refactoring stages to reduce the refactoring effort and, therefore, increase its adaption. Figure 14 shows the count of publications dealing with manual, semi-automatic, and automated refactoring. In fact, 30.95% of the papers deal with the automation of refactoring. Only 1.95% and 8.67% of the papers used manual and semi-automatic refactoring, respectively.

## 4.2 Artifacts affected by refactoring

As we mentioned before, refactoring is not limited to software code. In fact, it can be applied to any type of software artifacts (e.g., software architectures, database schema, models, user interfaces, and code). Figure 15 shows the percentage of refactoring publications per artifact. The evidence from this histogram shows that the most popular refactoring artifact is code (72.89%). Model refactoring has also received considerable attention, with a percentage of 59.25%. Graphical user interfaces (GUIs) and Database refactoring have received the least attention of all with a fraction of only 4.12% and 2.58%, respectively. This might be due to the fact that database refactoring is conceptually more difficult than code refactoring; code refactorings only need to maintain behavioral semantics while database refactorings also must maintain informational semantics. Also, GUI refactoring is very demanding, requiring the adoption of user interfaces architectural patterns from the early software design stages. Future research should explore database and user interface refactoring further as they are an indispensable part of today's software.

## 4.3 Refactoring objectives

Five paradigms have been identified from analyzing the primary studies: object-oriented designs, cloud computing, mobile apps, model-driven, and aspect-oriented. Object-oriented programming has gained popularity because it matches the way people actually think in the real world, structuring their code into meaningful objects with relationships that are obvious and intuitive. The increased popularity of the object-oriented paradigm has also increased the interest in object-oriented refactoring. This can be observed in figure 16 where more than 34% of the studies related to refactoring focus on object-oriented designs. Less than 5% of the papers investigated refactoring for cloud computing and mobile app development. For the refactoring objectives classification of the taxonomy, five subcategories are considered: external quality (e.g. correctness, usability, efficiency, reliability, etc.) , internal quality (e. g. maintainability, flexibility, portability, re-usability, readability etc.) , performance (e.g. response time, error rate, request rate, memory use, etc.), migration (e.g. Dispersion in the Class Hierarchy, number of referenced variables, number of assigned variables etc. ), security (e.g. time needed to resolve vulnerabilities, Number of viruses and spams blocked, Number of port probes, number of patches applied, Cost per defect, Attack surface etc.). Figure 17 is illustrating the reasons why people refactor their systems. Improving the internal quality takes up the largest portion (41.63%) followed by refactoring to improve the external quality (22.68%). Although security is a major concern for almost all systems, only 3.11% of the papers investigated refactorings for security reasons.

## 4.4 Refactoring techniques

Object-oriented programming languages have common traits/properties that facilitate the development of widely automated source code analysis and transformation tools. Many studies [25] have given sufficient proof that a refactoring tool can be built for almost any object-oriented language (Python, PHP, Java, and C++). Support for multiple languages in a refactoring tool is mentioned by [26]. Java is probably the most commercially important recent object-oriented language with an infrastructure that is designed to support analysis. It has generic parsing, tree building, prettyprinting, tree manipulation, source-to-source rewriting, attribute grammar evaluations, control, and data flow analysis. This explains the fact that 17.15% of refactoring studies (see figure 18) provided refactoring techniques and tools that support Java. At the same time, most of the other programming languages have a fraction of less than 1%. We classified the refactoring techniques into four main categories: data mining (e.g., Clustering, Classification, Decision trees, Association, Neural networks, etc.), search-based methods (e.g., Genetic algorithms, Hill climbing, Simulated annealing, Multi-objective evolutionary algorithms, etc.), formal methods (B-Method, the specification languages used in automated theorem proving, RAISE, the Z notation, SPARK Ada, etc.), and fuzzy logic. More than 25% of the papers use Search-based techniques to address refactoring problems (see figure 19). This can be explained by the fact that search-based approaches have been proven to be efficient at finding solutions for complex and labor-intensive tasks. With the growing complexity of software systems, there's an infinite amount of improvement/changes you can make to any piece of artifact. Exact algorithms are hard to use to solve the refactoring problem within an instance-dependent, finite run-time. That's why finding optimal refactoring solutions are sacrificed for the sake of getting perfect solutions in polynomial time using heuristic methods like search-based algorithms. Data mining techniques have also received significant attention (17.59%) as they are known to be efficient at discovering new information, such as unknown patterns or hidden relationships, from huge databases like, for our case, large code repositories.

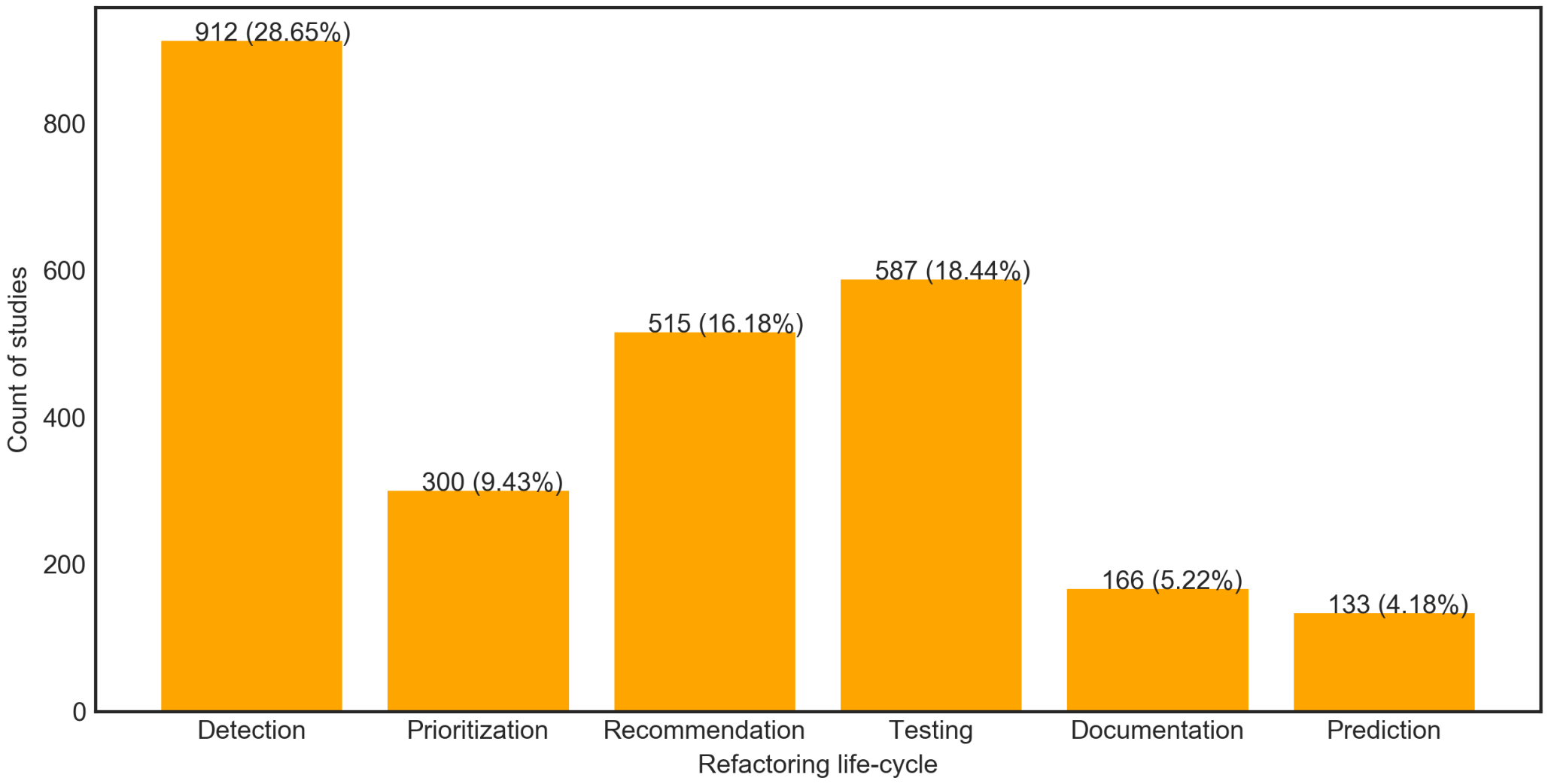

Fig. 13. Histogram illustrating the percentage of refactoring publications per refactoring life-cycle

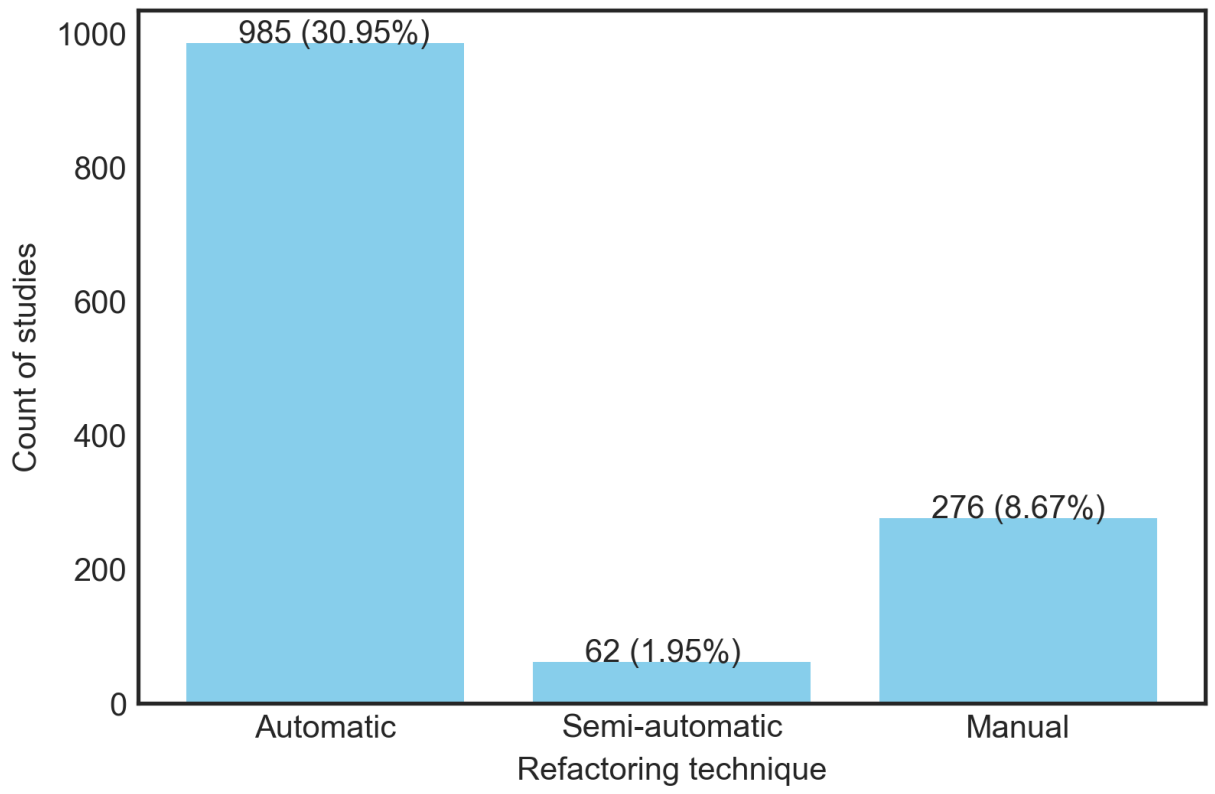

Fig. 14. Histogram illustrating the percentage of publications dealing with manual, semi-automatic and automated refactoring

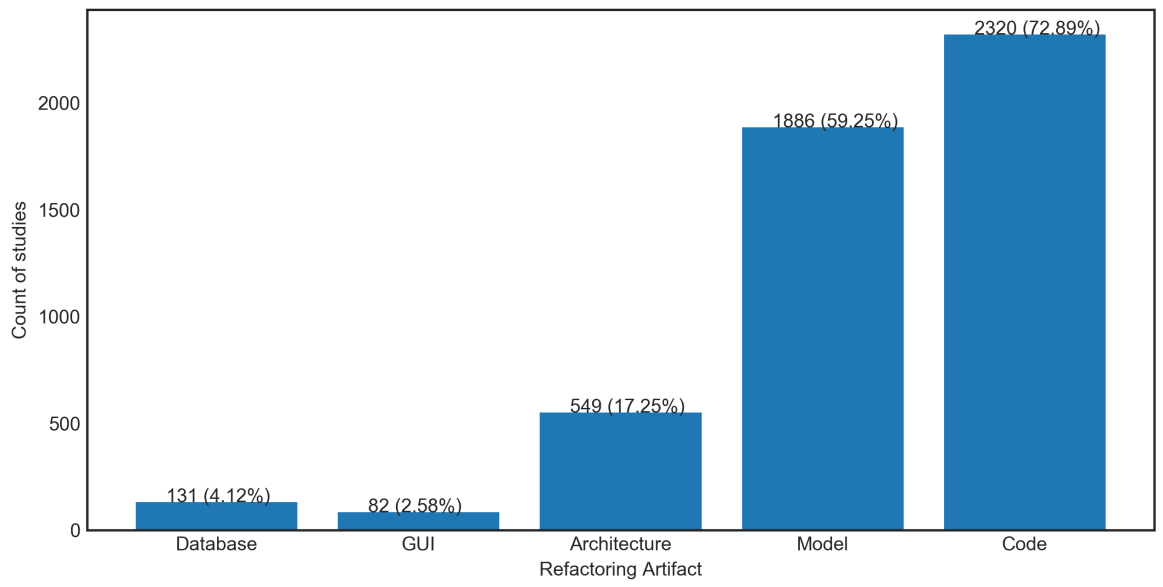

Fig. 15. histogram illustrating the count of refactoring publications per artifact

### 4.5 Refactoring evaluation

Open-source software systems are becoming increasingly important these days. 61.1% of the studies (see figure 20)

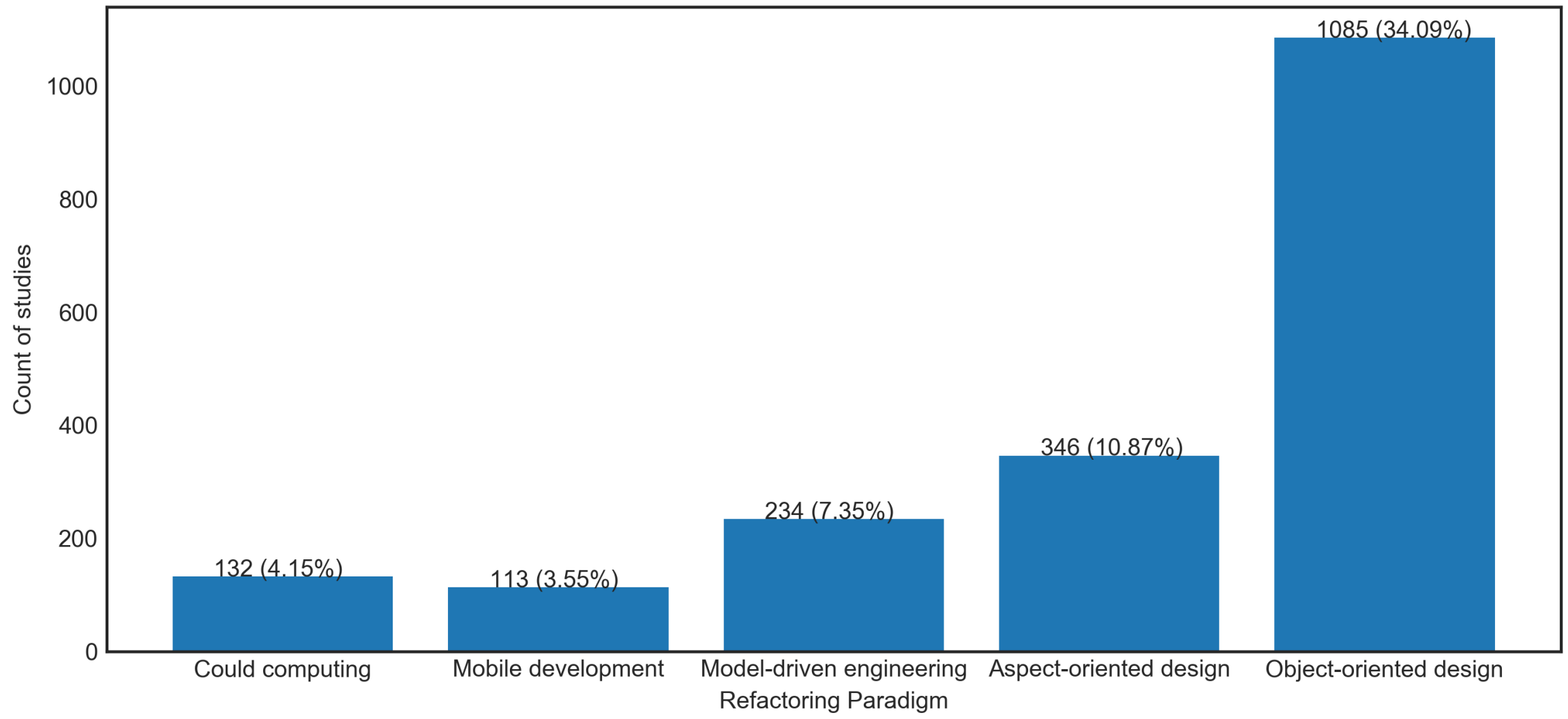

Fig. 16. Histogram illustrating the count of refactoring publications per paradigm

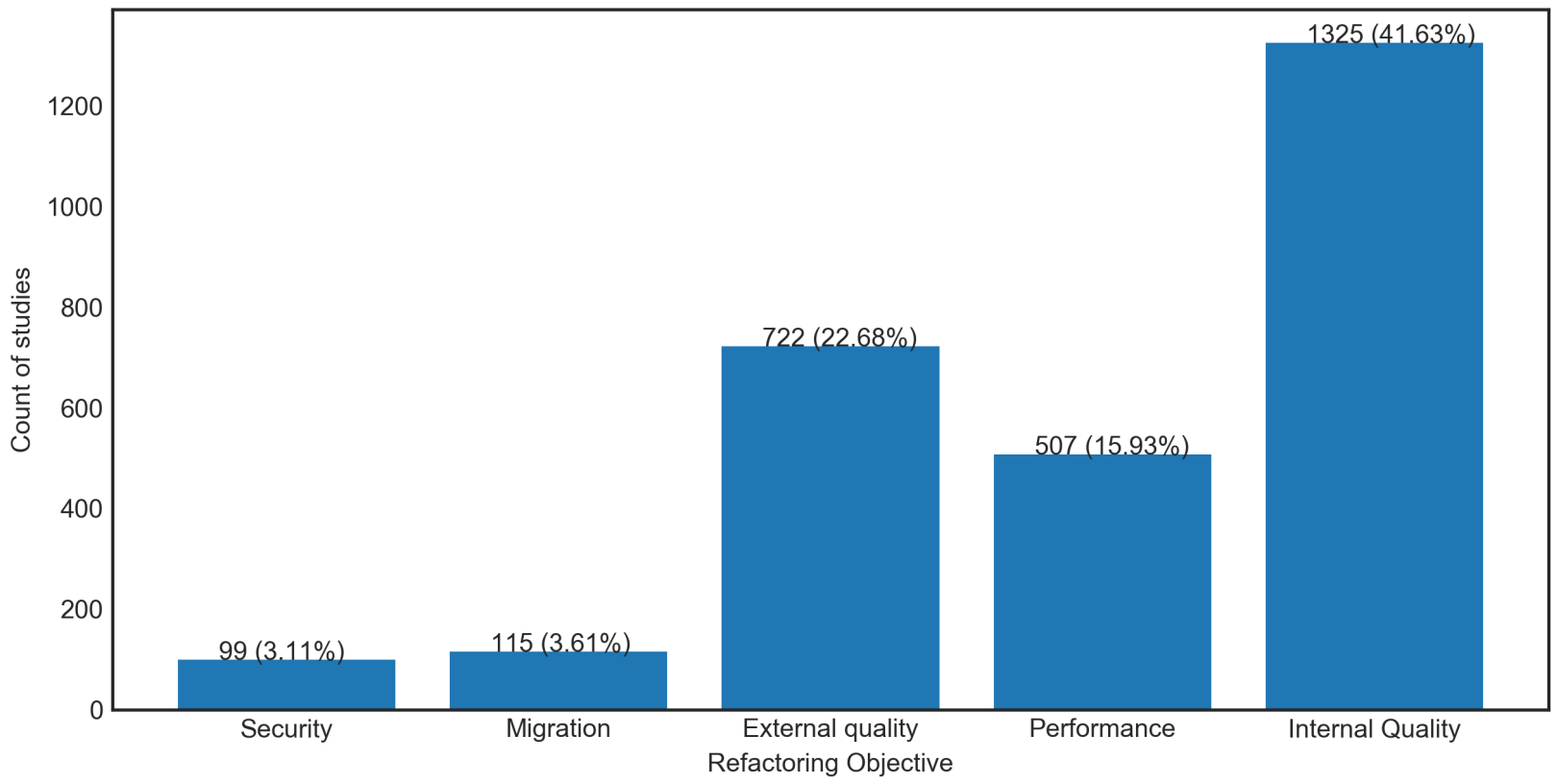

Fig. 17. Histogram illustrating the count of publications per refactoring objective

used open-source systems to validate their work compared to 38.9% of studies that validated their work on industrial projects. This result is expected because of the availability and accessibility of open source systems. However, open-source software is often developed with a different management style than the industrial ones. Thus, refactoring techniques and tools must be validated and checked for quality and reliability using industrial systems. More industrial collaborations are needed to bridge the gap between academic research and the industry's research needs, and therefore, produce groundbreaking research and innovation that solves complex real-world problems.

## 5 Conclusion

In this paper, we have conducted a systematic literature review on refactoring accompanied by meta-analysis to answer the defined research questions. After a comprehensive search that follows a systematic series of steps and assessing the quality of the studies, 3183 publications were identified. Based on these selected papers, we derived a taxonomy focused on five key aspects of Refactoring: refactoring life-cycle, artifacts affected by refactoring, refactoring objectives, refactoring techniques, and refactoring evaluation. Using this classification scheme, we analyzed the primary studies and presented the results in a way that enables researchers to relate their work to the current body of knowledge and identify future research directions. We also implemented a repository that helps researchers/practitioners collect and report papers about Refactoring. It also provides visualization charts and graphs that highlight the analysis results of our selected studies. This infrastructure will bridge the gap among the different refactoring communities and allow for more effortless knowledge transfer. To conclude, we believe that the results of our systematic review will help advance the refactoring research area. Since we expect this research area to continue to grow in the future, we hope that our repository and taxonomy will become useful in organizing, developing and judging new approaches.

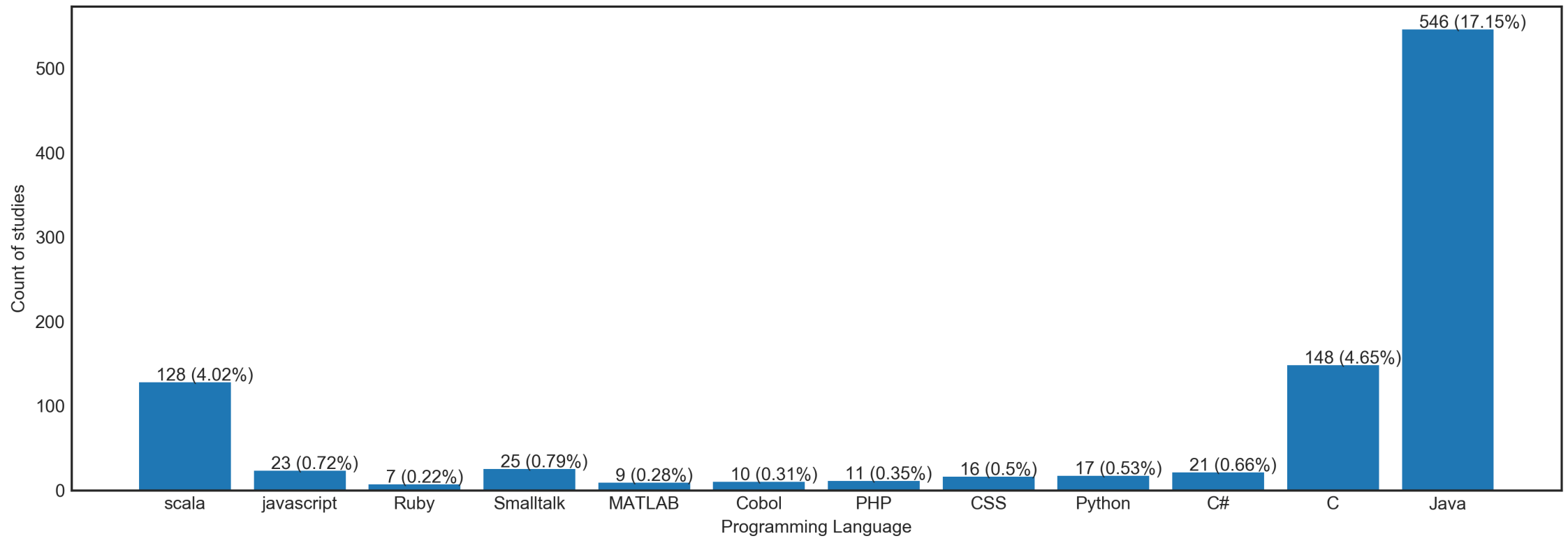

Fig. 18. histogram illustrating the count of refactoring publications per programming language

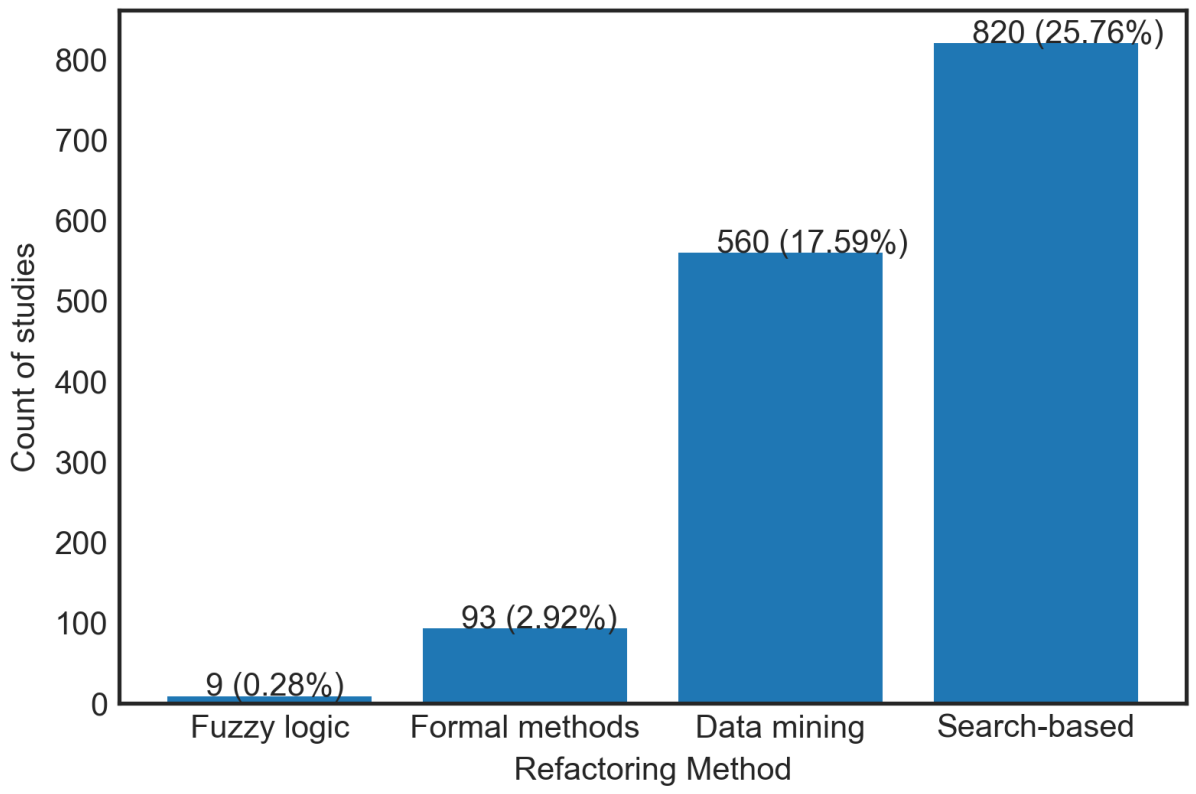

Fig. 19. histogram illustrating the count of refactoring publications per field

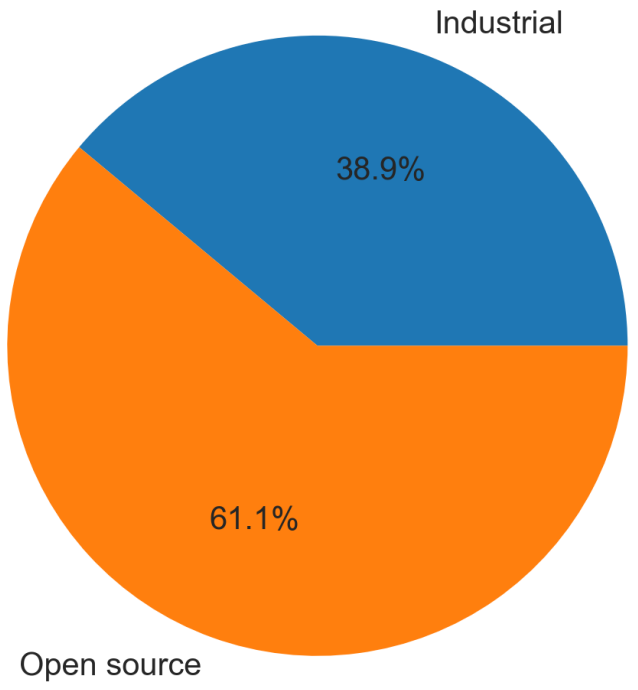

Fig. 20. Pie chart illustrating the percentage of publications in which the authors used industrial and/or open source systems in the validation step

## Primary Sources

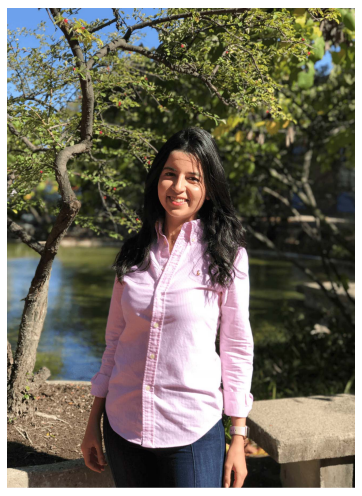

**Chaima Abid** is currently a PhD student in the intelligent Software Engineering group at the University of Michigan. Her PhD project is concerned with the application of intelligent search and machine learning in different areas such as web services, refactoring and security. Her current research interests are Search-Based Software Engineering, web services, refactoring, security, data analytics and software quality.

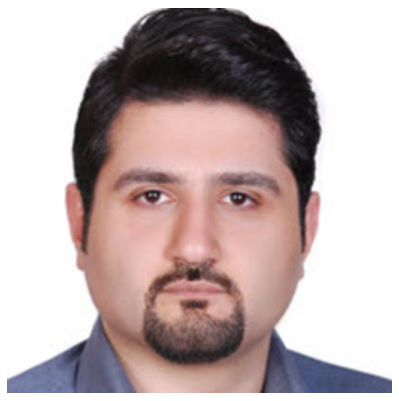

**Vahid Alizadeh** is currently a Ph.D. student in the intelligent Software Engineering group at the University of Michigan. His Ph.D. project is concerned with the application of intelligent search and machine learning in different software engineering areas such as refactoring, testing, and documentation. His current research interests are Search-Based Software Engineering, Refactoring, Artificial Intelligence, data analytics and software quality.

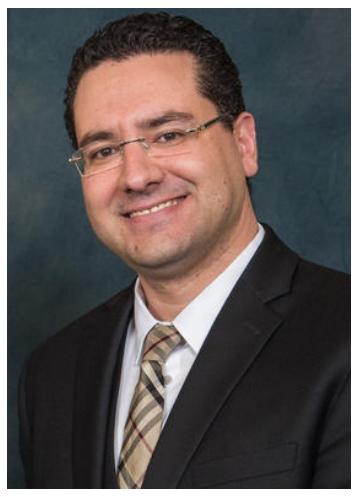

**Marouane Kessentini** is a recipient of the prestigious 2018 President of Tunisia distinguished research award, the University distinguished teaching award, the University distinguished digital education award, the College of Engineering and Computer Science distinguished research award, 4 best paper awards, and his AI-based software refactoring invention, licensed and deployed by industrial partners, is selected as one of the Top 8 inventions at the University of Michigan for 2018 (including the three campuses), among over 500 inventions, by the UM Technology Transfer Office. He is currently a tenured associate professor and leading a research group on Software Engineering Intelligence. Prior to joining UM in 2013, He received his Ph.D. from the University of Montreal in Canada in 2012. He received several grants from both industry and federal agencies and published over 110 papers in top journals and conferences. He has several collaborations with industry on the use of computational search, machine learning and evolutionary algorithms to address software engineering and services computing problems.

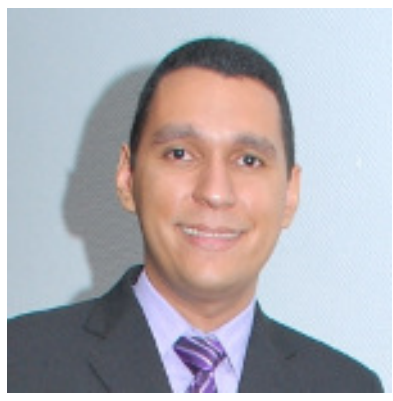

**Thiago do Nascimento Ferreira** is a Postdoctoral Researcher at University of Michigan-Dearborn under the supervision of Dr. Marouane Kessentini in the ISELab. He received my PhD Degree in Computer Science from the Federal University of Parana in 2019. His research mainly focuses on the use of Preference and Search Based Software Engineering to address several software engineering problems such as Software Testing and Software Refactoring.

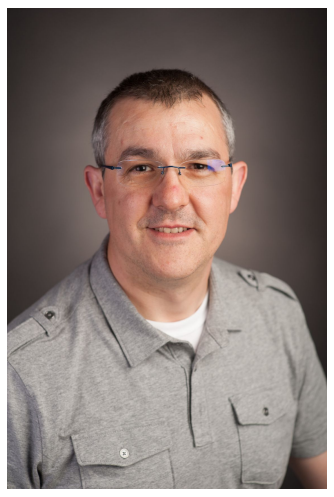

**Danny Dig** is an associate professor of computer science at the University of Colorado, and an adjunct professor at University of Illinois and Oregon State. He successfully pioneered interactive program transformations by opening the field of refactoring in cutting-edge domains including mobile, concurrency and parallelism, component-based, testing, and end-user programming. He earned his Ph.D. from the University of Illinois at Urbana-Champaign where his research won the best Ph.D. dissertation award, and the First Prize at the ACM Student Research Competition Grand Finals. He did a postdoc at MIT. He (co-)authored 50+ journal and conference papers that appeared in top places in SE/PL. According to Google Scholar his publications have been cited 4000+ times. His research was recognized with 8 best paper awards at the flagship and top conferences in SE, 4 award runner-ups, and 1 most influential paper award (N-10 years) at ICSME'15. He received the NSF CAREER award, the Google Faculty Research Award (twice), and the Microsoft Software Engineering Innovation Award (twice). He released 9 software systems, among them the world's first open-source refactoring tool. Some of the techniques he developed are shipping with the official release of the popular Eclipse, NetBeans, and Visual Studio development environments (of which Eclipse alone had more than 14M downloads in 2014).